\documentclass[aps,prb,reprint,twocolumn,superscriptaddress,letterpaper,floatfix,preprintnumbers,showpacs]{revtex4-1}
\usepackage{amsfonts}
\usepackage{amsmath}
\usepackage{amssymb}
\usepackage{bm}
\usepackage{color}
\usepackage{color,graphicx}
\usepackage{dcolumn}	% Align Table  columns on decimal point
\usepackage{gensymb}
\usepackage{graphicx}	% Include Figure files
\usepackage[hyperfootnotes,hyperindex,hyperfigures]{hyperref}
\usepackage{latexsym}
\usepackage{mathcomp}
\usepackage{textcomp}
\usepackage{verbatim}
\usepackage{dcolumn}
\usepackage{ulem}
\usepackage{url}

\providecommand{\degr}{$^{\circ}$}
\providecommand{\STO}{SrTiO$_{3}$}
\providecommand{\LTO}{LaTiO$_{3}$}
\providecommand{\LAO}{LaAlO$_{3}$}
\providecommand{\tio}{TiO$_6$~}

\definecolor{red}{rgb}{0.8,0,0.2}

\def\beeq{\begin{equation}}
\def\eneq{\end{equation}}
\def\beeqa{\begin{eqnarray}}
\def\eneqa{\end{eqnarray}}

\definecolor{adobe}{rgb}{.8,.6,.5}
\definecolor{mygreen}{rgb}{0.0,0.6,0.0}
\definecolor{blue2}{rgb}{0.0,0.0,0.8}
\definecolor{brown}{rgb}{0.6,0.3,0.0}
\definecolor{forest}{rgb}{0.0,0.4,0.0}
\definecolor{grass}{rgb}{0.0,0.55,0.25}
\definecolor{grass2}{rgb}{0.0,0.6,0.25}
\definecolor{gray}{rgb}{0.4,0.4,0.4}
\definecolor{grayish}{rgb}{0.2,0.2,0.4}
\definecolor{khaki}{rgb}{0.9,0.9,0.7}
\definecolor{lightteal}{rgb}{0.0,0.6,0.6}
\definecolor{lightyellow}{rgb}{1.0,1.0,0.5}
\definecolor{maroon}{rgb}{0.7,0.1,0.2}
\definecolor{navy}{rgb}{0.0,0.1,0.7}
\definecolor{olive}{rgb}{0.4,0.4,0.0}
\definecolor{orange}{rgb}{0.9,0.45,0.0}
\definecolor{peach}{rgb}{1.0,.8,.7}
\definecolor{purple}{rgb}{0.4,0,0.55}
\definecolor{teal}{rgb}{0.0,0.5,0.4}
\definecolor{turq}{rgb}{0.3,0.6,0.9}
\definecolor{violet}{rgb}{0.75,0,0.75}
\hypersetup{colorlinks=true, linkcolor=navy, citecolor=blue, urlcolor=blue, filecolor=red, anchorcolor=blue}

\begin{document}
\DeclareGraphicsExtensions{.ps,.pdf,.eps}

\preprint{UPDATED: {\color{maroon} \today}}

%: TITLE OF PAPER +++++++++++++++++++++++++++++++++++++++++++++++++++++++++++
\title{Structural Phase Transitions of the Metal Oxide Perovskites SrTiO$_{3}$, LaAlO$_{3}$ and LaTiO$_{3}$ Studied with a Screened Hybrid Functional}

%: AUTHOR LIST ++++++++++++++++++++++++++++++++++++++++++++++++++++++++++++++++
\author{\firstname{Fedwa} \surname{El-Mellouhi}}
   \email{fadwa.el\_mellouhi@qatar.tamu.edu}
     \affiliation{Department of Chemistry, Texas A\&M at Qatar, Texas A\&M Engineering Building, Education City, Doha, Qatar}

\author{\firstname{Edward N.} \surname{Brothers}}
   \email{ed.brothers@qatar.tamu.edu}
     \affiliation{Department of Chemistry, Texas A\&M at Qatar, Texas A\&M Engineering Building, Education City, Doha, Qatar}

\author{\firstname{Melissa J.} \surname{Lucero}}
  \affiliation{Department of Chemistry, Rice University, Houston, Texas 77005-1892}

\author{\firstname{Ireneusz W.} \surname{Bulik}}
  \affiliation{Department of Chemistry, Rice University, Houston, Texas 77005-1892}

\author{Gustavo E. Scuseria}
  \affiliation{Department of Chemistry, Rice University, Houston, Texas 77005-1892}
  \affiliation{Department of Physics and Astrononmy, Rice University, Houston, Texas 77005-1892}
 \affiliation{Chemistry Department, Faculty of Science, King Abdulaziz University, Jeddah 21589, Saudi Arabia }
	%\email{Electronic mail: guscus@rice.edu}
	%\homepage{http://python.rice.edu/~guscus/}

%: ABSTRACT ++++++++++++++++++++++++++++++++++++++++++++++++++++++++++++++++  
\begin{abstract}
We have investigated the structural phase transitions of the transition metal oxide perovskites SrTiO$_{3}$, LaAlO$_{3}$ and LaTiO$_{3}$ using the screened hybrid density functional of Heyd, Scuseria and Ernzerhof (HSE06). 
We show that HSE06-computed lattice parameters, octahedral tilts and rotations, as well as  electronic properties, are significantly improved over semilocal functionals. 
We predict the crystal field splitting ($\Delta_{CF}$) resulting from the structural  phase transition in  SrTiO$_{3}$ and LaAlO$_{3}$ to be 3 meV and 10 meV, respectively, in excellent agreement with experimental results. HSE06   identifies correctly  LaTiO$_{3}$ in the magnetic sates as a Mott insulator. Also, it predicts that the GdFeO$_{3}$-type distortion in  non-magnetic LaTiO$_{3}$ will induce a  large $\Delta_{CF}$ of 410 meV. This large crystal-field splitting associated with the large magnetic moment found in the G-type antiferromagnetic state suggest that  LaTiO$_{3}$ has an  induced orbital order, which is confirmed by the visualisation of the highest occupied orbitals. These results strongly indicate that HSE06 is capable of efficiently and accurately modeling perovskite oxides, and promises to efficiently capture the physics at their heterointerfaces. 
\end{abstract}

\pacs{71.15.Mb,% Density functional theory, local density approximation , gradient and other corrections.
 71.15.Ap,% Basis sets (LCAO, plane-wave, APW, etc.) and related methodology (scattering methods, ASA, linearized methods, etc.)
 77.80., %Ferroelectricity and antiferroelectricity
 77.84.} %  Dielectric, piezoelectric, ferroelectric, and antiferroelectric materials

\maketitle
\clearpage

 % ====
\section{Introduction}
 % ====
Heterointerfaces between metal oxides like \STO~(STO), \LAO~(LAO)~and \LTO~(LTO) show considerable promise as components in all-oxide electronics because they exhibit unique properties unobserved in the corresponding isolated parent compounds. 
In their bulk phase, \STO~and \LAO~are non-magnetic wide-bandgap materials, but when assembled into superlattices, interesting properties such as high $T_{c}$ superconductivity, magnetism, ferroelectricity and colossal magnetoresistance are observed.  
Despite recent experimental and theoretical activities, the field of oxide interfaces remains full of surprises and unsolved problems.\cite{Chakhalian:2012,Hwang:2012}

 A number of  heterointerface behaviors have been explained using density functional theory (DFT), but several authors (see Refs.~\onlinecite{Pentcheva:2010, Zubko:2011} and references therein) have shown that more elaborate methods like dynamical mean field theory\cite{Kotliar:2006,Held:2006,Vollhardt:2012} or GW-based approaches\cite{Onida:2002} are needed to successfully model interfaces of strongly correlated electronic systems and predict the direction of charge transfer. 
While many-body techniques in combination with DFT methods are sufficient for this task, they remain computationally demanding, particularly when it is necessary to account for the important structural relaxations and phase transitions occurring at interfaces. 

Theory's predictive power can be enhanced by considering the  properties of the two (or more) parent materials in the bulk phase, followed by mapping to obtain heterointerface properties.\cite{Trimarchi:2011,Franceschetti:1999} 
To successfully model the perovskites considered herein, good band gaps and accurate structural/geometric properties beyond reasonable lattice parameters are required. 
For example, the \tio rotation and tilt angles are likely responsible for the metal-insulator transition in oxide superlattices\cite{Rondinelli:2010} and necessitate accurate theoretical structures. 

Controlling the octahedral rotations relies on good geometric values of TiO$_6$ and is considered by some to be the key to designing functional metal oxides.\cite{Benedek:2011} 
Further, the \tio~rotation and/or tilts affect the electronic properties of  both bulk metal oxides and their superlattices.
For example, the  degeneracy of the $t_{2g}$ states in (LAO)$_5$/(LTO)$_n$/(LAO)$_5$~superlattices is lifted by the crystal field induced by the \tio octahedral rotation in \LTO~layer.\cite{Seo:2010}
 Consequently, accurate determination of  crystal field splittings (CFS) induced by the phase transition  in bulk metal oxides is also important to the Ti-$3d$ orbital reconstruction at the heterointerfaces.\cite{Aguado:2011}  

Recently, Jalan \textit{et al.}\cite{Jalan:2011} found that stress also exerts a pronounced influence on the electron mobility in  STO, with moderate strains resulting  in  a greater than 300\% increased mobilities.
Good theoretical estimation of strain in metal oxides is thus recommended and computational methods approaching the experimental strain are preferred.
This is the case for the recent HSE06 calculations of Janotti \textit{et al.},\cite{Janotti:2011} which confirmed the experimental findings and showed that strain in STO affects seriously the electronic effective masses  and the conductivity.

A survey of literature (see Ref.~\onlinecite{El-Mellouhi:2011} and references therein) presented in Table~\ref{tab:comparison} below, summarizes the main semilocal and hybrid functional trends for bulk \STO, a widely-used substrate for the heterointerface superlattices of interest here.  Methods underestimating the band gaps, like the Local Spin Density Approximation\cite{LSDA} (LSDA), the generalized gradient approximation functional of Perdew, Burke and Ernzerhof\cite{Perdew:1996,Perdew:1997} (PBE), and the PBEsol\cite{Staroverov:2003,Staroverov:2004} re-parameterization for solids must be avoided because they lead to an overestimation of the two-dimensional band at the heterointerfaces, and tend to overestimate the octahedral rotation angle in \STO. 

In addition, LSDA and PBE also fail to describe the antiferromagnetic insulating character in strongly-correlated systems %like \LTO, predicting them to be metallic conductors.~\cite{Rivero:2009}
 
%%%
\begin{table*}[!hbt]
\caption{Trends observed in the performance of DFT and post-DFT calculations applied to bulk \STO~regardless of the basis sets used (see  Ref.~\onlinecite{El-Mellouhi:2011}). Hybrids account for both global (B3PW,\cite{Zhukovskii:2009,Heifets:2006,Piskunov:2004} B3LYP\cite{Piskunov:2004} and B1-WC.\cite{Bilc:2008}) and screened (HSE and HISS) hybrid functionals.}
\begin{ruledtabular}
\begin{tabular}{lllllll}
\label{tab:comparison}
Functional		&LSDA		&PBEsol		&PBE		&DFT+U		&Hybrids\\
\hline
\\
Lattice parameter	&Too Low	&Good		&Too Big	&Too Low	&Good\\
\\
Band gaps		&Too Small 	&Too Small 	&Too Small 	& Good 		& Good \\
				&~-40\%	&~-40\%	&~-40\%	& 			&$\pm$6\%\\
\tio rotation angle	&Too Big	&Too Big	&Too Big	&Too Big	&Good\\
				&Up to 8\degr	&5\degr &4\degr &4-5\degr &~1-2\degr\\	
\end{tabular}
\end{ruledtabular}
\end{table*}
%%%

DFT$+U$ methods can open the band gap to the experimental value by empirical tuning of the value of the Coulomb on-site ($U$) and exchange ($J$) interaction parameters, but at the cost of incorporating parameters that are external to theory and are material-dependent. 
While DFT$+U$ can correctly describe the insulating character of \LTO~in the ground state,\cite{Ahn:2006} geometric issues like the rotation angle overestimation problems in STO, which originate from the parent functionals LSDA and PBE, do not improve by adding the $+U$ correction.  
Despite these well-known limitations,  DFT+$U$ has been the method of choice for the oxide superlattice calculations, mainly because the corrected gaps approach the experimental band gaps of  \STO, \LAO~and \LTO~bulk phases with a minimal additional cost compared to regular DFT calculations, especially for superlattices.\cite{Schwingenschlogl:2008,Gemming:2006} 

Very important progress has been made in furthering our understanding of metal oxides superlattices using DFT+$U$; however, one value of $U$ will not produce the correct band gap for differing bulk metal oxides, which poses a problem for heterostructure systems like LAO/STO, LTO/STO, LTO/LAO, etc.

In fact, most DFT+$U$ calculations assume that the value of $U$ appropriate for the substrate is also valid to describe the grown material, as well as the heterointerface,\cite{Ahn:2006,Rondinelli:2010,Pardo:2010} although in some instances the $U$ correction is applied to the bulk only, while the grown material is treated at the regular DFT level.\cite{Lee:2008} 
(Even more sophisticated methods use different values of $U$ and $J$ for the Ti-3$d$ and La-4$f$ states; see for example Ref.~\onlinecite{Ong:2011}). 
Another problem recently gaining notoriety is  the fact  that DFT+$U$  predicts exaggerated octahedral distortions in the fully-relaxed superlattices,\cite{Ong:2011} which seriously affects the electronic structure and the properties of the 2-D electron gas at the interfaces. 
All of these shortcomings indicate the need for a universally applicable functional to better describe the electronic structure of thin films and their interfaces.

Electronic structure calculations on bulk \STO~using the global hybrid functionals B3PW,\cite{Becke:1993} B3LYP,\cite{B3LYP,Lee:1988} and B1-WC\cite{Bilc08} were performed by several groups,\cite{Zhukovskii:2009,Eglitis:2008,Heifets:2006} showing improved  band gaps compared to semilocal DFT. 
	% --> There is no Alexandrov:200x in the bib files I have, so I have removed the citation for the time-being -- MJL
For example, band gap deviations from experiment were only 6-7\% with B3PW,  while structural properties and order parameters of the antiferrodistortive (AFD) phase  (see Table~\ref{tab:comparison}) were also closer to experiment. 
Nevertheless, the use of global hybrids in studying heterointerfaces remains limited due to their high computational cost.

Screened hybrid density functionals like that of Heyd, Scuseria, and Ernzerhof\cite{HSE06,HSE06e} (HSE/HSE06) and the middle-range screened hybrid functional of Henderson, Izmaylov, Scuseria and Savin (HISS)\cite{HISS2} are excellent candidates for this task due to their accuracy and much lower computational cost compared to regular hybrids.\cite{Lucero:2012,Henderson:2011,Janesko:2009,Henderson:2008my} 
	% --> Added 2 more Scuseria reviews: Ben's & Tom's -- per GES
Screened hybrid functionals yield \STO~structural properties for both phases\cite{El-Mellouhi:2011,Wahl08} in very good agreement with experiment,\cite{Yamada:2010} especially if used with large (but still computationally tractable) basis sets.
The best results arise from the HSE/TZVP combination.\cite{El-Mellouhi:2011}
	% --> Moved Table II, the \LAO~Lattice parameters table, to the LAO section.

Previous calculations on iron~\cite{Paier:2006} and transition metal monoxides~\cite{Marsman:2008} using HSE03\cite{HSE03} indicate that the magnetic moment, exchange splitting and band width in metals are overestimated relative to experiment. 
Subsequent tests on strongly-correlated magnetic materials have been carried out by Rivero \textit{et al.}\cite{Rivero:2009} using the short-range screened hybrid HSE06 and the long-range screened hybrid LC-$\omega$PBE.\cite{LCwPBE}
They demonstrated that both range-separated  hybrids give a good quantitative description of the electronic structure of strongly-correlated systems, with magnetic coupling constants that are larger than experiment, yet improved compared to the results of the global hybrid, B3LYP.

A more recent work form Philipps and Peralta\cite{Phillips:2011} involving transition metal molecular complexes suggests that HSE06 is a promising alternative for the evaluation of exchange couplings in extended systems, again because of reasonable accuracy at reduced computational expense.
The present paper is a continuation of earlier work\cite{El-Mellouhi:2011} with \STO: herein, we assess the accuracy of  HSE06 for calculating the crystal field splitting in \STO, the geometries of \LAO~in its different phases, as well as the structural and magnetic properties of the strongly-correlated system, \LTO.

 % ===
\section{Computational Methods} 
 % ===
All calculations presented in this paper were performed using a development version of the {\sc Gaussian} suite of programs,\cite{gdv} with the periodic boundary condition
(PBC)\cite{Kudin:2000} code used throughout.   
Unless otherwise noted, crystal structures were  downloaded as CIF files from the ICSD.\cite{ICSD} (See Ref.~\onlinecite{ICSD} for the specific ID numbers.)
	% --> Need to fill in blanks for ICSD collection numbers in the notes part of ICSD reference e.g., there are multiple 		 structures for the Eitel citation, so it is not clear which structure was used.

The  Def2-~\cite{Weigend:2005} series of Gaussian basis sets were optimized for use in bulk \LTO~and \LAO~calculations, following the procedure described in Ref.~\onlinecite{El-Mellouhi:2011} for bulk \STO.
We use the notation TZVP and SZVP  to differentiate these optimized PBC basis sets from the molecular  Def2-TZVP and Def2-SZVP basis sets. 
The functionals applied in this work include the LSDA\cite{LSDA} (SVWN5),\cite{SVWN5} PBE, and HSE06. 
LSDA and PBE calculations were utilized to assess the quality of the Gaussian basis sets via comparison with planewave calculations from the literature. 

Most numerical settings in {\sc gaussian} were left at the default values, \textit {e.g.}, geometry optimization settings, integral cut-offs, $k$-point meshes and SCF convergence thresholds. 
For \LAO,~the reciprocal space integration used a 12$\times$12$\times$12 $k$-point  mesh for the cubic unit cells of 5 atoms, while for the larger rhombohedral  supercell of 30 atoms,  the default  $k$-point  mesh of 8$\times$8$\times$4 was found to be sufficient.
Because we performed optimizations of both atomic positions and lattice parameters, the SCF convergence threshold was set to {\sc{tight}}, or 10$^{-8}$ atomic units. The fully relaxed structures can be obtained from  the Cambridge Structural Database. \footnote{ CCDC 900592 - 900596 contains the supplementary crystallographic data for this paper. These data can be obtained free of charge from The Cambridge Crystallographic Data Centre via \url{www.ccdc.cam.ac.uk/data_request/cif }}

Due to the metallic nature of \LTO, a mesh of 24$\times$24$\times$18 for the $k$-point sampling was required in order to ensure stable convergence.\cite{Paier:2006}  
For the sake of computational efficiency, all calculations on LTO were carried out using the SZVP basis set.

 % ===
\section{Results and Discussion}
 % ===
 % -----------
\subsection{\STO}
 % -----------
Under ambient conditions, bulk \STO~crystallizes in a cubic perovskite structure,  subsequently undergoing a second order phase transition at $T_c$=105 K to a tetragonal structure with slightly rotated  oxygens around the $z$-axis, known as the antiferrodistortive (AFD) phase. 
The phase transition of STO is governed by two order parameters.
The primary order parameter is the rotation angle of the TiO$_6$ octahedra ($\theta$) that can reach up to  2.1\degree~at 4.2~K.\cite{Unoki:1967}
The second order parameter measures the tetragonality of the unit cell, as defined by the ratio of $c/a$, and can be as large as 1.0009 upon cooling to 10~K.\cite{Heidemann:1973}

The finer details of the conduction band splitting induced by the AFD phase transition in STO are still under debate,\cite{Bistritzer:2011,Chang:2010, Uwe:1985,Mattheiss:1972} with discrepancies between theory and experiment arising from differences in the sample dopant or defect concentrations: the \tio~rotation angle may be enhanced or reduced depending on the doping conditions and this affects directly the rotation of oxygen atoms in the $xy$-plane when STO distorts into the AFD phase. 

We have shown\cite{El-Mellouhi:2011} in modeling an idealized STO (with no defects, doping, strain or surface effects) that HSE06 provides band gaps and phase transition order parameters in excellent agreement with experiment.  We found a  $c/a$=1.0012 corresponding  to a +0.12\% tensile strain along the [001] direction or the $z$-axis.
However, we did not address the crystal field splitting ($\Delta_{CF}$) of the conduction band minimum (CBM) induced by the phase transition. 
As is evident from the enlargement of the region around the calculated CBM  at the $\Gamma$ point 
 the crystal field causes the 3-fold degenerate band to split into a doubly-degenerate band and a single band 3 meV higher in energy. 
\begin{figure*}[!htb]
\includegraphics[width=0.3\textwidth]{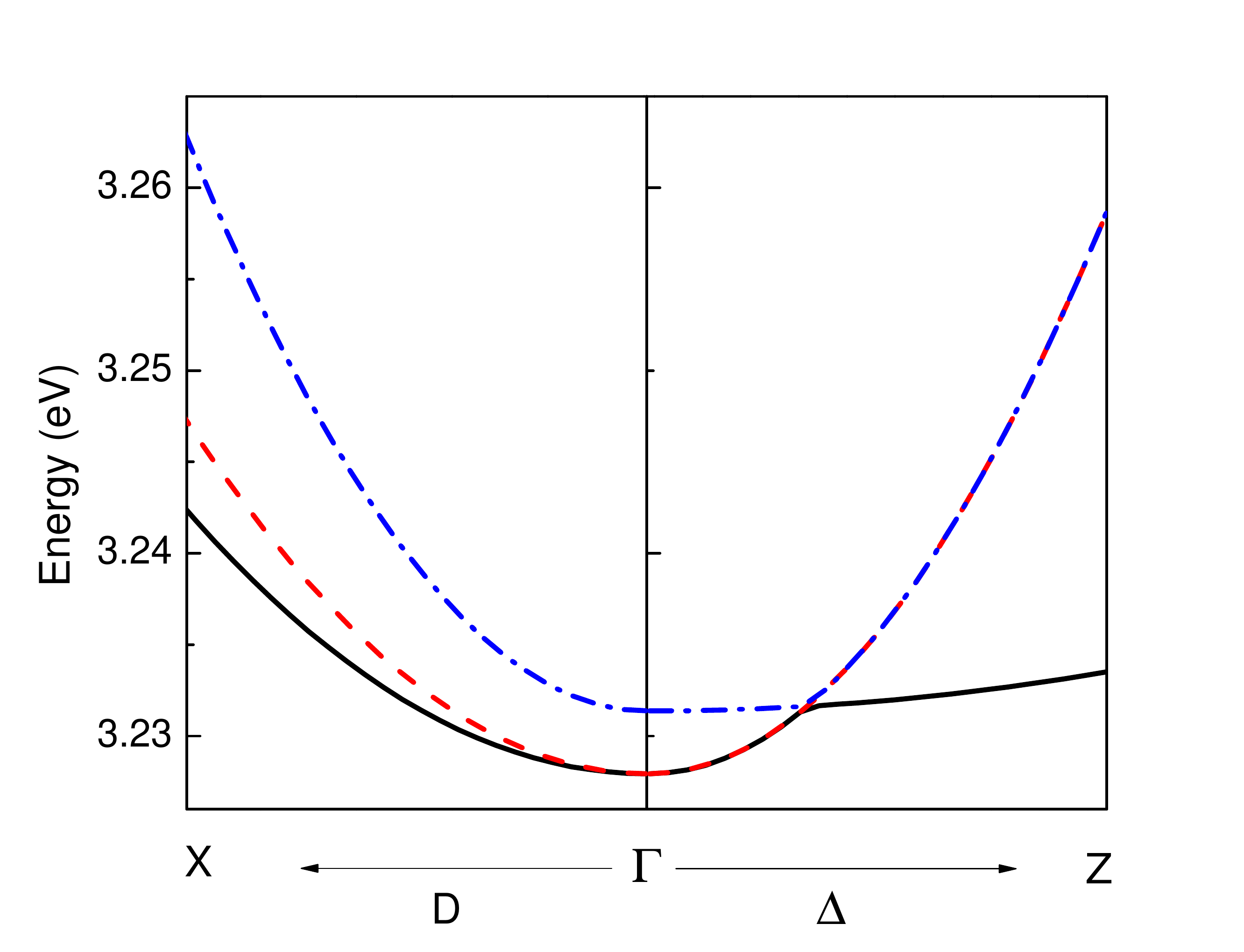}
\includegraphics[width=0.3\textwidth]{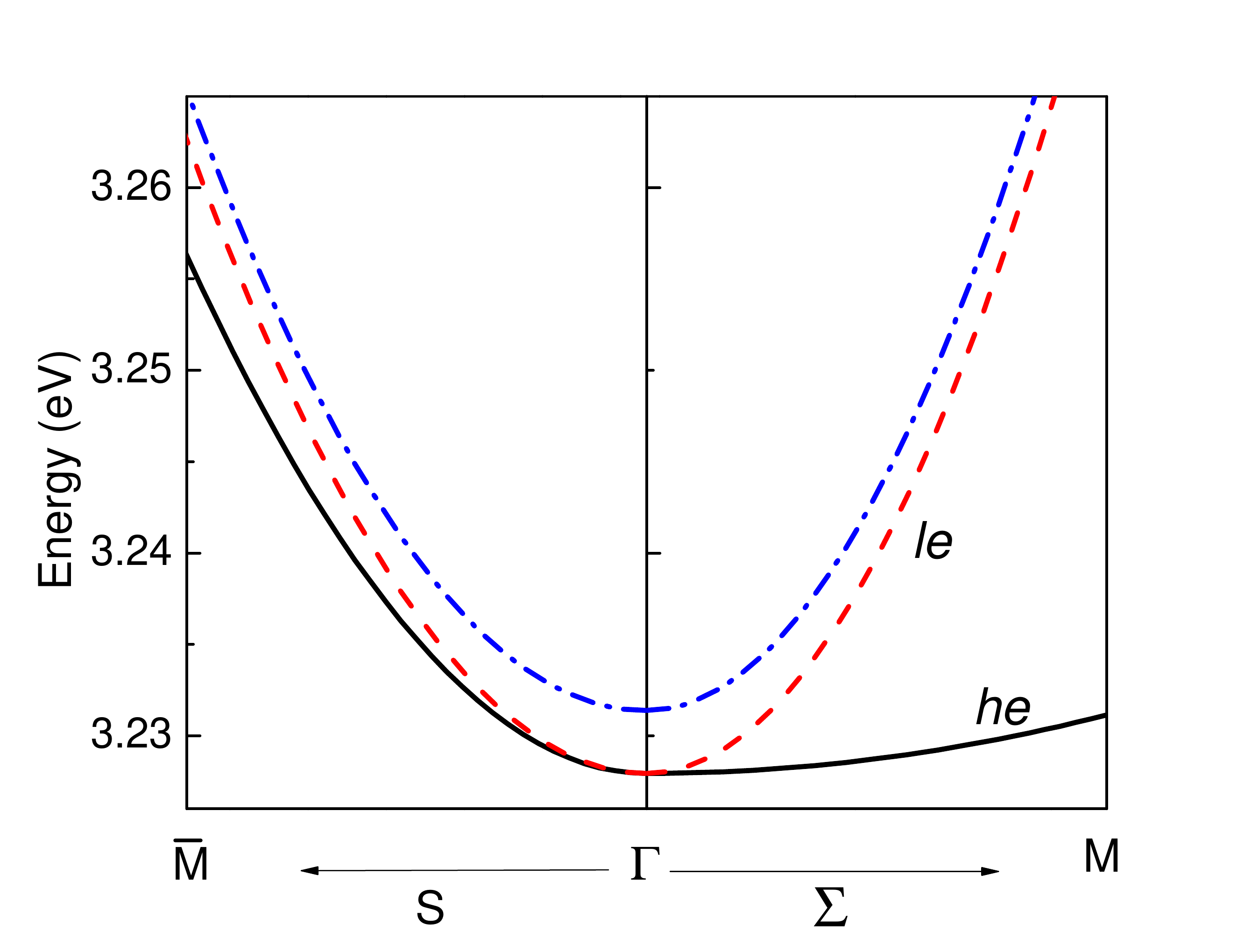} % generated via gnuplot & cropped
\includegraphics[width=0.3\textwidth]{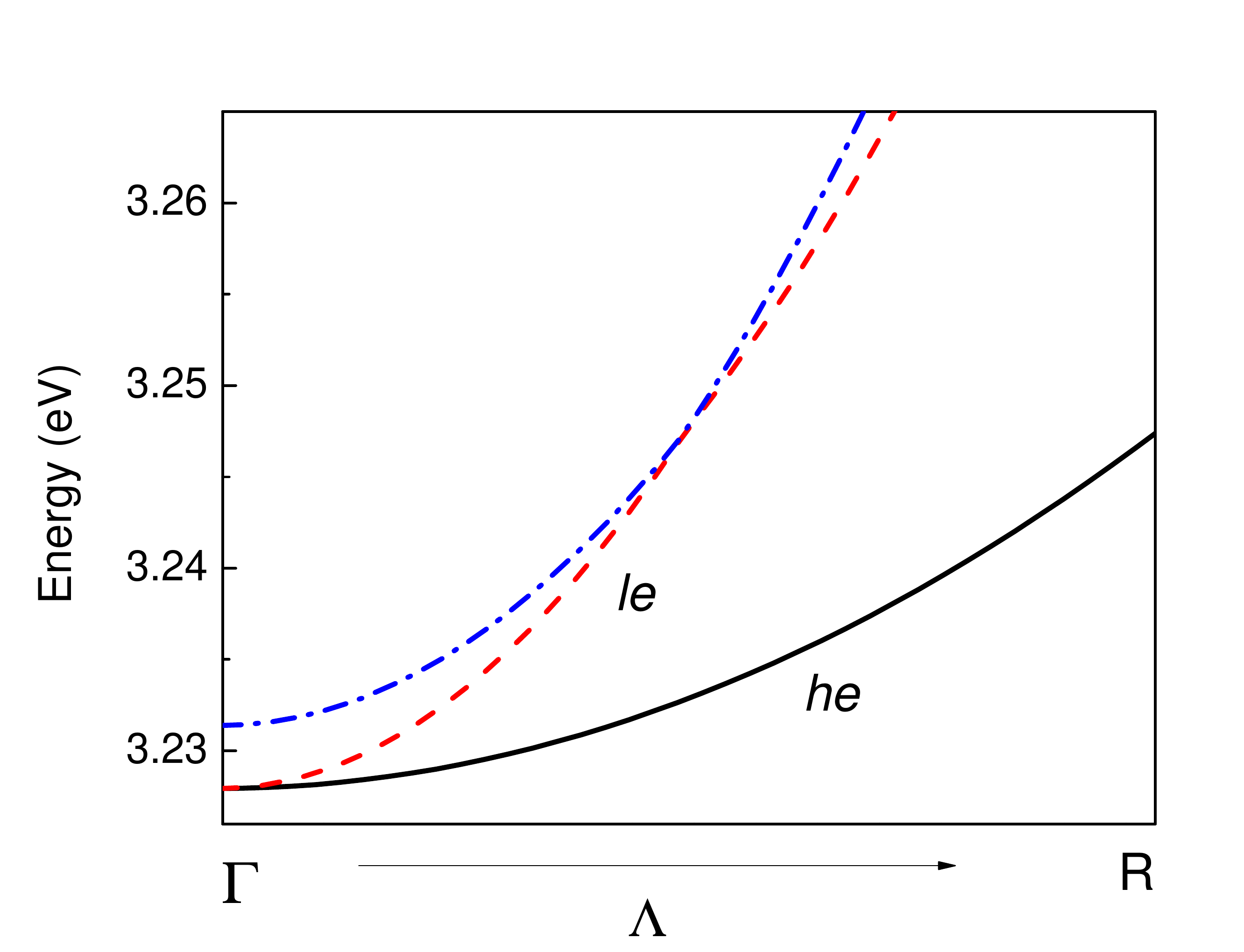}
\caption{\label{fig:STO}\
(Color online)  Enlargement of the region around calculated conduction band minimum (CBM) at the $\Gamma$ point showing the  splitting in \STO caused by the  AFD phase transition.  
The high symmetry directions are those of the body centered tetragonal lattice  [100] ($D$), [001] ($\Delta$), [101] (S), [110] ($\Sigma$), [111] ($\Lambda$)   and $Z$ =(00$\frac{1}{2}$) $\bar M$ =($\frac{1}{2}0\frac{1}{2}$). }
\end{figure*}

Further discussion arise from  figure~\ref{fig:STO} by zooming at 10\%  from  Brillouin zone center along the high symmetry directions, [100] ($\Delta$), [001] ($D$), [101] (S), [110] ($\Sigma$) and  [111] ($\Lambda$) in the body centered tetragonal Brillouin zone. Along $\Delta$, $D$, and S,  we observe a  qualitative behavior very similar to that predicted in an earlier augmented plane wave (APW) calculation without spin-orbit (SO)  coupling.\cite{Mattheiss:1972} However, differences arise along $\Sigma$ and $\Lambda$ directions where we observe that the lowest doubly degenerate band splits into a  heavy electron ($he$) and light electron ($le$) bands as we move farther from the $\Gamma$ point.
The differences might come from the model used in the APW calculation by Mattheiss neglecting both the nearest and second-nearest neighbor oxygen-oxygen $2p$ interactions and the interaction between the oxygen-$2s$ and the titanium $t_{2g}$ orbitals in the cubic state.

 Quantitatively, the APW calculations predict $\Delta_{CF}^{APW}$=20.7 meV  which is much greater than the $\Delta_{CF}^{HSE06}=3$ meV we obtain in the present work. This is also the case for the spin-orbit splitting  for cubic STO $\Delta_{SO}^{APW}$=83 meV, if compared with  $\Delta_{SO}^{HSE06}$=28 meV computed using the VASP planewave code  by Janotti et {\it al.}.\cite{Janotti:2011}

Janotti et {\it al.} studied with HSE06 the effect of $\pm$1\% strain on the CBM focusing on the splittings and the electron effective masses taking into account the spin-orbit effects.  They also found that the rotation of \tio octahedra  around the $z$-axis without including   the tetragonal distortion ($c>a$) lead to a splitting of the doubly degenerate band of  4~meV with minor changes on the effective masses.  Going back to our results in figure~\ref{fig:STO},  one concludes that the  small +0.12\% tensile strain along the [001] direction arising from the tetragonal distortion might not contribute significantly to the CBM splitting but plays a role on  the important differences  in the electron effective masses we observe along the high symmetry directions. For example, the curvature of the CBM is higher (i.e. lower effective mass or light electrons-{\it le}) along $S$ and $\Delta$  directions --parallel to the strain-- indicating  higher electron mobilities in those directions. This result  is in line with the enhanced electron mobility  Janotti et {\it al.} concluded for  \STO~ subject to +1\% tensile strains along the (001) direction.  However,  further investigation   is still needed to draw a final conclusion since  the electron effective masses and the order of the bands change upon the inclusion of SO effects.~\cite{Janotti:2011}

The combined  HSE06 results, $\Delta_{CF}^{HSE06}$ and $\Delta_{SO}^{HSE06}$ show an excellent  \textit{quantitative} agreement with experimental data obtained from single domain Shubnikov-de Haas oscillation Raman spectroscopy measurements,\cite{Uwe:1985} namely: $\Delta_{CF}^{exp}$=2 meV and  $\Delta_{SO}^{exp}$=18 meV. However, they  do not support the picture proposed by the recent angle resolved photoemission spectroscopy (ARPES) experiment of Chang \textit{et al.},\cite{Chang:2010} which indicates that the single band is of lower energy, with a doubly-degenerate band 25 meV higher. This disagreement may arise from the fact that STO samples used in this experiment were doped with oxygen vacancies (V$_O$), which are known to introduce a shallow defect level in the band gap of STO.\cite{Liu:2011} 
Finally, using HSE06 and the TZVP basis set to model STO relaxed in the AFD phase, we found that the crystal field splits the CBM at the gamma point  by  3 meV, leaving it doubly degenerate in the absence of SO coupling. 
% ------------
\subsection{LaAlO$_{3}$}
% ------------
Bulk \LAO~undergoes a phase transition from  the simple cubic structure to a rhombohedral-central hexagonal structure with space group $R\bar3C$ at temperatures below 813~K.\cite{Howard:2000,Hayward:2005,Lehnert:2000}   
The two order parameters for this phase transition are the tilt of the AlO$_6$ octahedra ($\Theta$), which can reach a maximum value of  5.7\degree at 4.2K,\cite{Hayward:2005} and $\tau$, a measure of spontaneous strain, defined as $\tau=c/a-\sqrt{6}$, with values of up to -0.008.
The calculated lattice parameters relative to experiment for both phases of bulk LAO are underestimated with LSDA, overestimated with PBE and in-between for HSE06 (Table~\ref{tab:lattice}). 
This is  well-known behavior for LSDA and PBE (see \textit{e.g.} Ref.~\onlinecite{Mori:2008} and
references therein), and has been observed in a large number of semiconducting materials, as well as in our recent study on \STO.\cite{El-Mellouhi:2011}
%
 %%%
 \begin{table*}[!htb]
\caption{Calculated lattice parameters of LAO as well as  phase transition order parameters: the  octahedral tilt angle ($\Theta$)  and the spontaneous strain  ($\tau=c/a-\sqrt{6}$).  Values with a reference number come from previous experimental and/or planewave calculations.}
\begin{ruledtabular}
\begin{tabular}{lllllllll}
\label{tab:lattice}

	&\multicolumn{5}{c}{Rhombohedral}	&Cubic\\

	&a (\AA)					&b (\AA)				&c (\AA)			&$\Theta$(\degree) &$\tau$		&a=b=c (\AA)\\
\hline
Experiment\footnotemark[1]	&5.365 	&5.365  	&13.111 &5.7&$-$0.008	&3.810\footnotemark[2]\\
LSDA	&5.341, 5.290	\footnotemark[5]  &5.339	&13.020	&4.1, 6.1\footnotemark[5]	&$-$0.012	&3.760, 3.750\footnotemark[3]\\
PBE	&5.448, 5.370\footnotemark[4] 	&5.444, 5.370\footnotemark[4] &13.250,13.138\footnotemark[4] &4.0 	&$-$0.017	&3.830 \\
HSE06	&5.398	&5.393	&13.160	&3.1	&$-$0.011	&3.800
\end{tabular}
\end{ruledtabular}

\footnotetext[1]{At 4.2 K from Ref~\onlinecite{Hayward:2005}}
\footnotetext[2]{Ref~\onlinecite{Cwik:2003}}
\footnotetext[3]{Ref~\onlinecite{Knizhnik:2005}}
\footnotetext[4]{Ref~\onlinecite{Luo:2008}}
\footnotetext[5]{Ref~\onlinecite{Seo:2011}}
\end{table*}

In the low temperature rhombohedral phase, the LSDA and PBE lattice parameters tend to be higher than previous planewave calculations.\cite{Knizhnik:2005, Luo:2008,Seo:2011} 
The octahedral tilt angle for both LSDA and PBE is ca. 4\degree, which is 30\% smaller than the maximum tilts measured experimentally. 
The SZVP basis set produces tilt angles 30\% lower than those observed by planewave calculations\cite{Seo:2011} which is very similar to the behavior we observed in the case of STO: we attribute this to the use of localized basis sets.\cite{El-Mellouhi:2011}
HSE06 yields lattice parameters for the cubic phase in excellent agreement with experiment, with a deviation of only 0.25\%, and improved rhombohedral lattice parameters deviating  by 0.5\% compared to 1.5\% with PBE. 
Improvement of the same magnitude was also obtained using the recently introduced variational pseudo self-interaction correction approach(VPSIC$_0$).\cite{Filippetti:2011}

The calculated  spontaneous strain ($\tau$) is also significantly closer to experiment for HSE06 and LSDA than PBE. 
For all functionals considered here, the calculated lattice mismatch between cubic LAO and STO agree with the  2\% value reported in the experiment, indicating that the basis set we used will not introduce any additional strain in LAO/STO superlattices beyond those inherently present in the experimentally measured ones.

Turning now to the electronic properties of LAO, our computed band gaps are summarized in Table~\ref{tab:gap}. 
%
%%% 
\begin{table}[!htb]
\caption{Calculated band gaps of \LAO~using the SZVP basis set compared to experimental values. Data from this work are listed first, followed by data from previous planewave simulations and citations. }
\begin{ruledtabular}
\begin{tabular}{lllll}
\label{tab:gap}
	&Rhombohedral	&\multicolumn{2}{c}{Cubic}\\
Band gap(eV)			&Direct			&Indirect &Direct \\
\hline
Experiment	&5.60\footnotemark[1] 	&$-$	&$-$\\
LSDA	&3.75, 3.87\footnotemark[5] &3.25, 3.3\footnotemark[2] & 3.46\\
PBE	&3.98, 3.95\footnotemark[3] &3.27, 3.1\footnotemark[6] &3.54\\
HSE06	&5.24	&4.77	&5.04\\
\end{tabular}
\end{ruledtabular}
\footnotetext[1]{Ref~\onlinecite{Lim:2002}}
\footnotetext[2]{Ref~\onlinecite{Knizhnik:2005}}
\footnotetext[3]{Ref~\onlinecite{Luo:2008}}
\footnotetext[4]{Ref~\onlinecite{Luo:2009}}
\footnotetext[5]{Ref~\onlinecite{Seo:2011}}
\footnotetext[6]{Ref~\onlinecite{Xiong:2008}}
\end{table}
%%%
Comparisons with experimental band gaps are relative to the room temperature rhombohedral phases since, to our knowledge, no experimental measurements of the cubic phase band gap exist.

In general, band gaps calculated using the LSDA and PBE  are underestimated in average by $\sim$30\% or $\sim$1.8~eV compared to experiment, with PBE doing better than the LSDA.
This is consistent with earlier calculations\cite{Luo:2008, Knizhnik:2005} and indicates that the basis set does not negatively impact electronic structure properties in a significant manner, a behavior we observed in the STO case as well.\cite{El-Mellouhi:2011} (Please see Table~\ref{tab:lattice} for comparisons with previous calculations.) 
As expected, HSE06 provides a much better estimate of the band gaps, even when using the smaller SZVP basis sets, with deviations from experiment not exceeding  0.36 eV or 6\% for the rhombohedral phase. 
This error is larger than the 1\% deviation  observed in  the \STO~case,  but is well-within the 0.3-0.4 eV range for non-ionic bulk semiconductors.\cite{Lucero:2012,Matsushita:2011,Henderson:2011,Janesko:2009} 

\begin{figure*}[!htb]
\includegraphics[width=0.45\textwidth]{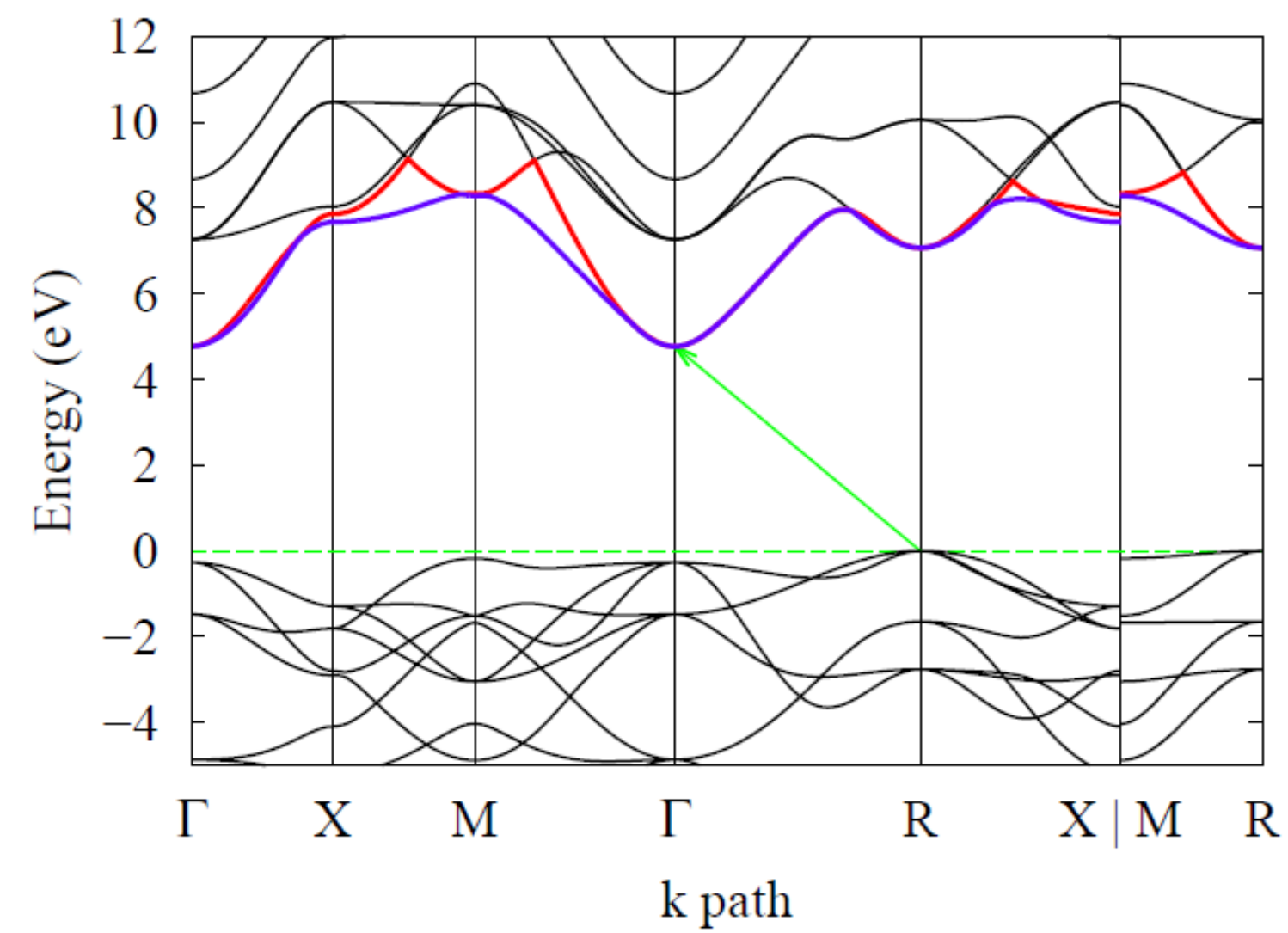}
\includegraphics[width=0.45\textwidth]{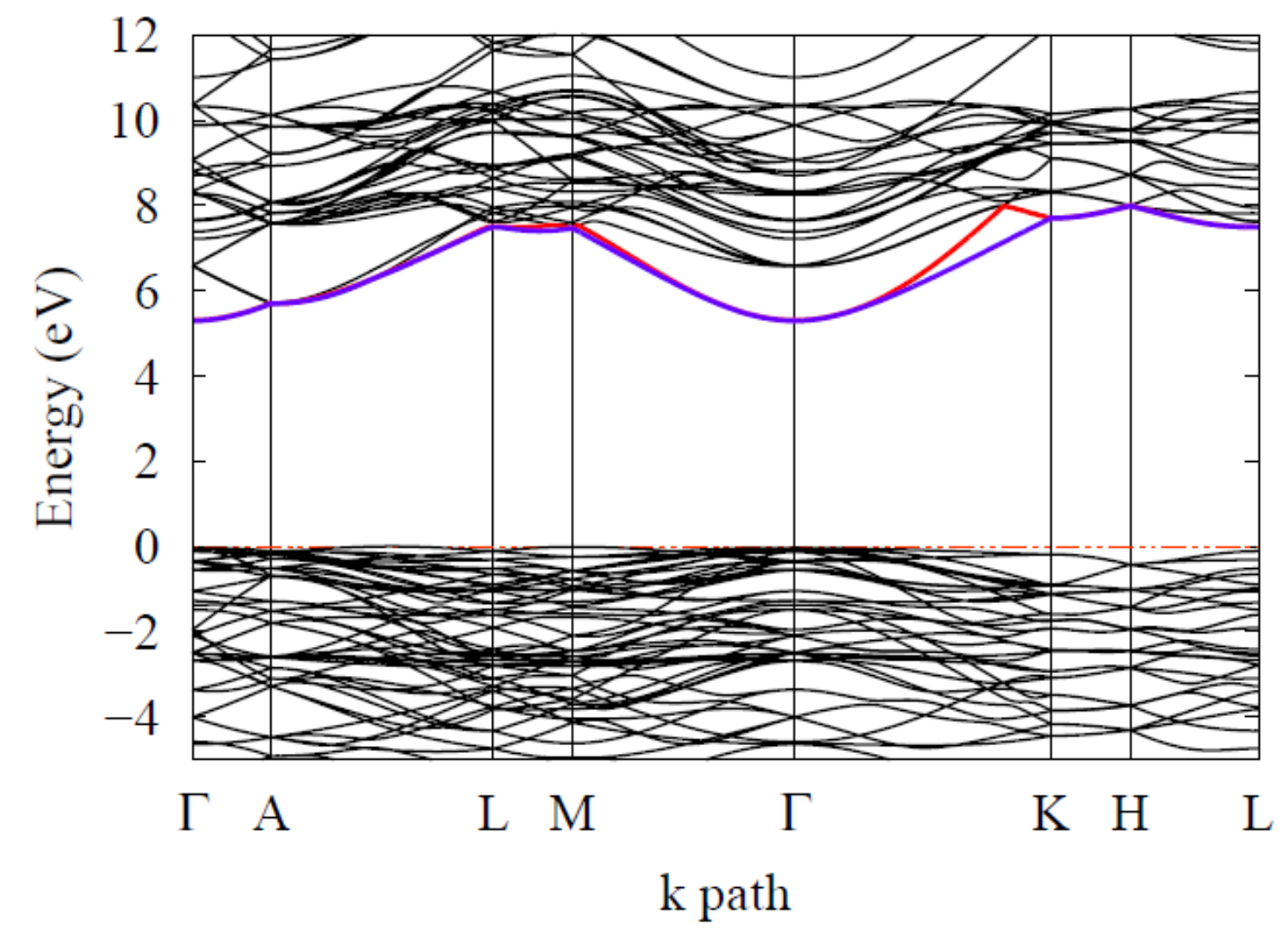}
\caption{\label{fig:LAO_band}(Color online) Band structure of \LAO~ calculated using HSE06 for the high temperature cubic phase (\textbf{Left}) and the  low temperature rhombohedral phase (\textbf{Right}).  The dashed line depicts the Fermi level at the valence band maximum (VBM).} 
\end{figure*}

For the cubic phase, HSE06  produces an indirect  band gap of 4.7~eV,  which agrees well with the calculated gap of 4.4 eV obtained using the screened exchange method\cite{Xiong:2008,Robertson:2006} (sX) and planewave basis sets and the  4.61~eV gap obtained using the VPSIC$_0$ method.\cite{Filippetti:2011}
The HSE06 direct gap is 5.0 eV, which lies between the  VPSIC$_0$~\cite{Filippetti:2011} value of 4.89~eV and the GW-corrected LSDA band gap at the experimental lattice constants\cite{Luo:2009} of 5.68~eV.
The quality of the cubic phase results is encouraging for future heterostructure and defect state calculations because LAO films grown on  Si substrates, using a few STO layers as a template, usually adopt the cubic structure.\cite{Reiner:2009}
The cubic and rhombohedral phase band structures as calculated by HSE06 for \LAO~are shown in Figure~\ref{fig:LAO_band}. 
They look very different because  the Brillouin zone sampling and the number of atoms in each simulation supercell  are different, but as we will demonstrate with the projected densities of states (PDOS), only small changes in the spectrum  occur at the phase transition.

Starting from the cubic phase, the band  gap is indirect (R$\rightarrow\Gamma$) with a doubly degenerate  CBM and a triply degenerate valence band maximum, VBM. 
A close look at our computed CBM in the rhombohedral phase  also reveals that the phase transition induces a lift of degeneracy at the CBM and a splitting at the $\Gamma$ point with $\Delta_{CF}$=10 meV. 
The band gap increase and  $\Delta_{CF}$  for LAO are much higher than the values we reported for \STO; we attribute this to  the fact that six of six oxygens are experiencing significant rigid octahedral tilts leading to changes in the La-O-La angles from the ideal 180\degree~to 173.8\degree, compared to a smaller rotation resulting from a 4/6 oxygen ratio in the STO case. 

By aligning the VBM of both phases (see the right of Figure~\ref{fig:LAO_band}) we observe that the octahedral tilt leads to a shift of the CBM to higher energies, thus increasing the band gap by  500 meV. 
There is also experimental evidence that the crystal field caused by AlO$_6$ tilts in \LAO~induces splitting of the CBM; but no quantitative values were reported.\cite{Hayward:2005}

\begin{figure}[!htb]
\includegraphics[width=0.45\textwidth]{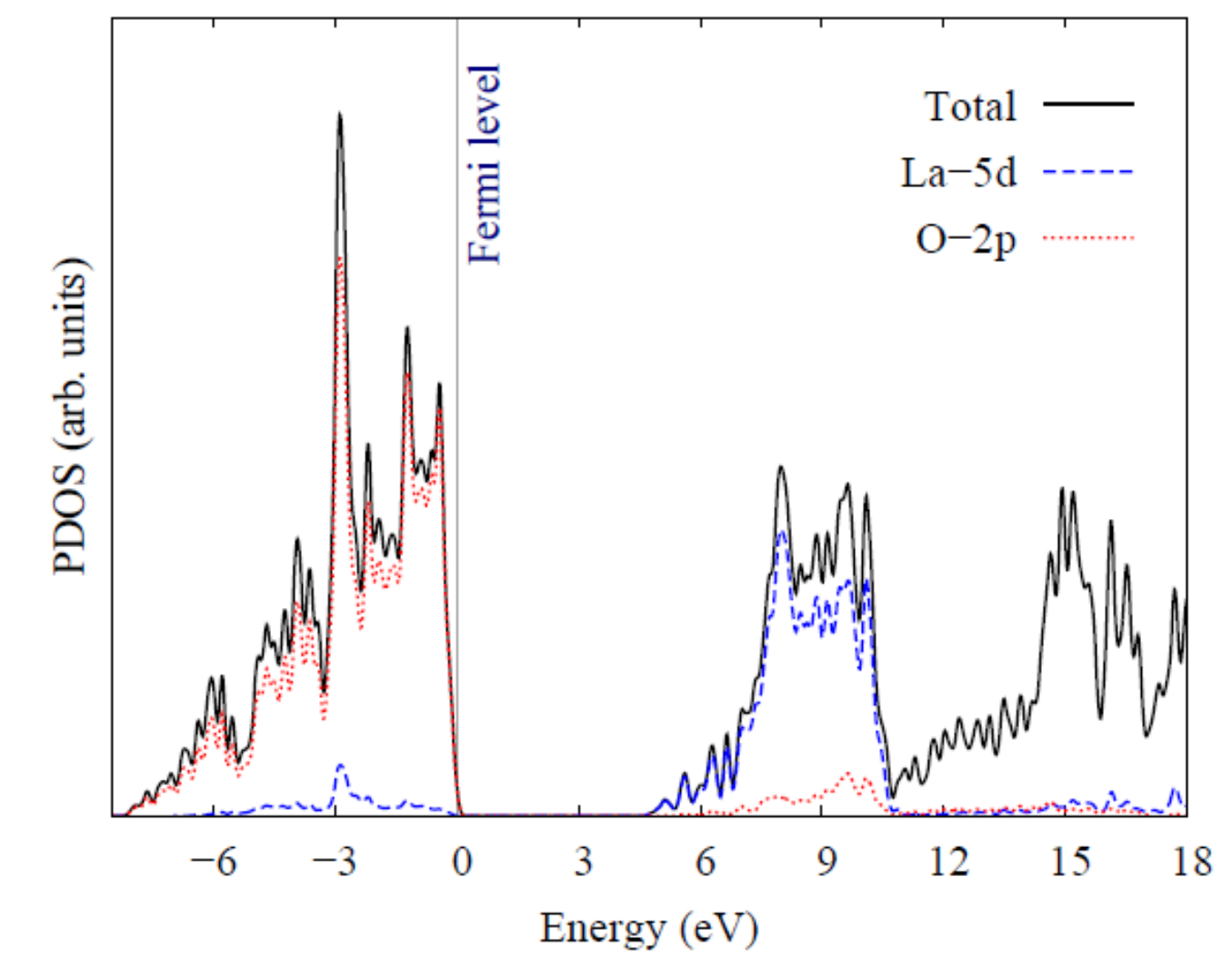}
\includegraphics[width=0.45\textwidth]{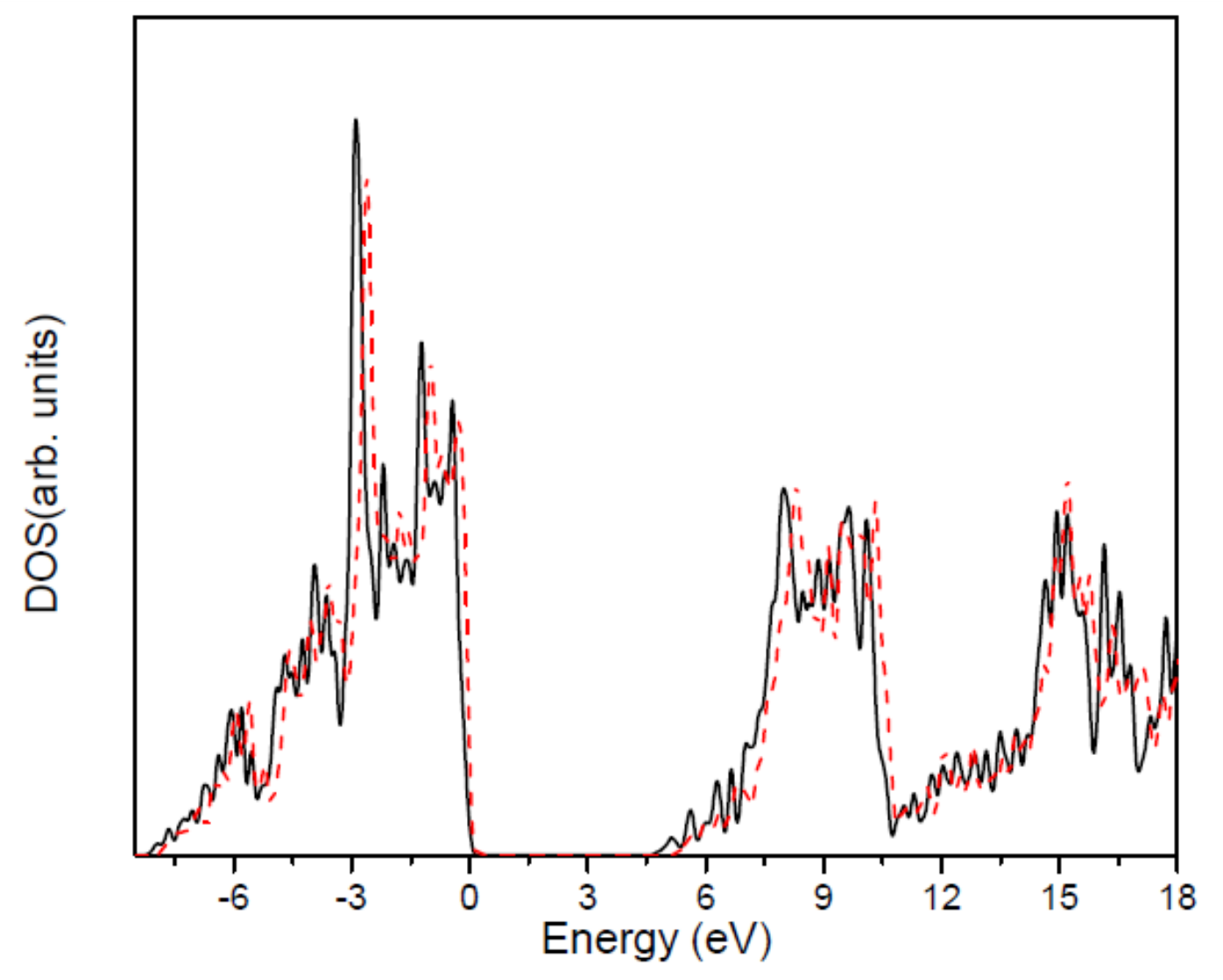}
\caption{\label{fig:LAO_pdos}(Color online) \textbf{Top}: Projected electronic densities  of states (PDOS) for \LAO~ calculated using HSE06 in the cubic phase. \textbf{Bottom}: Changes in the total density of states upon the phase transition form cubic (solid line) to rhombohedral (dashed line) mainly characterized by a shift of the conduction band states to higher energies.} 
\end{figure}

By looking at the PDOS (Figure~\ref{fig:LAO_pdos}) we observe that the VBM is strictly dominated by O-2p states, while La-5d states dominate the CBM. 
Unlike the \STO~case, no significant intermixing between the O-2p and La-5d states is observed, indicating that the bonding has predominately ionic  character.  
This picture is not affected by the phase transition, and no signs of orbital intermixing is observed in the rhombohedral phase PDOS as well, where  peaks conserve their height and shape (not shown). 

In summary, by applying HSE06 to LAO we lose some agreement with experiment on the octahedral tilts but we gain better precision than semilocal functionals in the band gaps, lattice parameters and strain altogether.
In other words, overall HSE06 provides a more physically accurate picture.  

% ------------
\subsection{\LTO}
\begin{figure}[!htb]	
\includegraphics[width=0.5\textwidth]{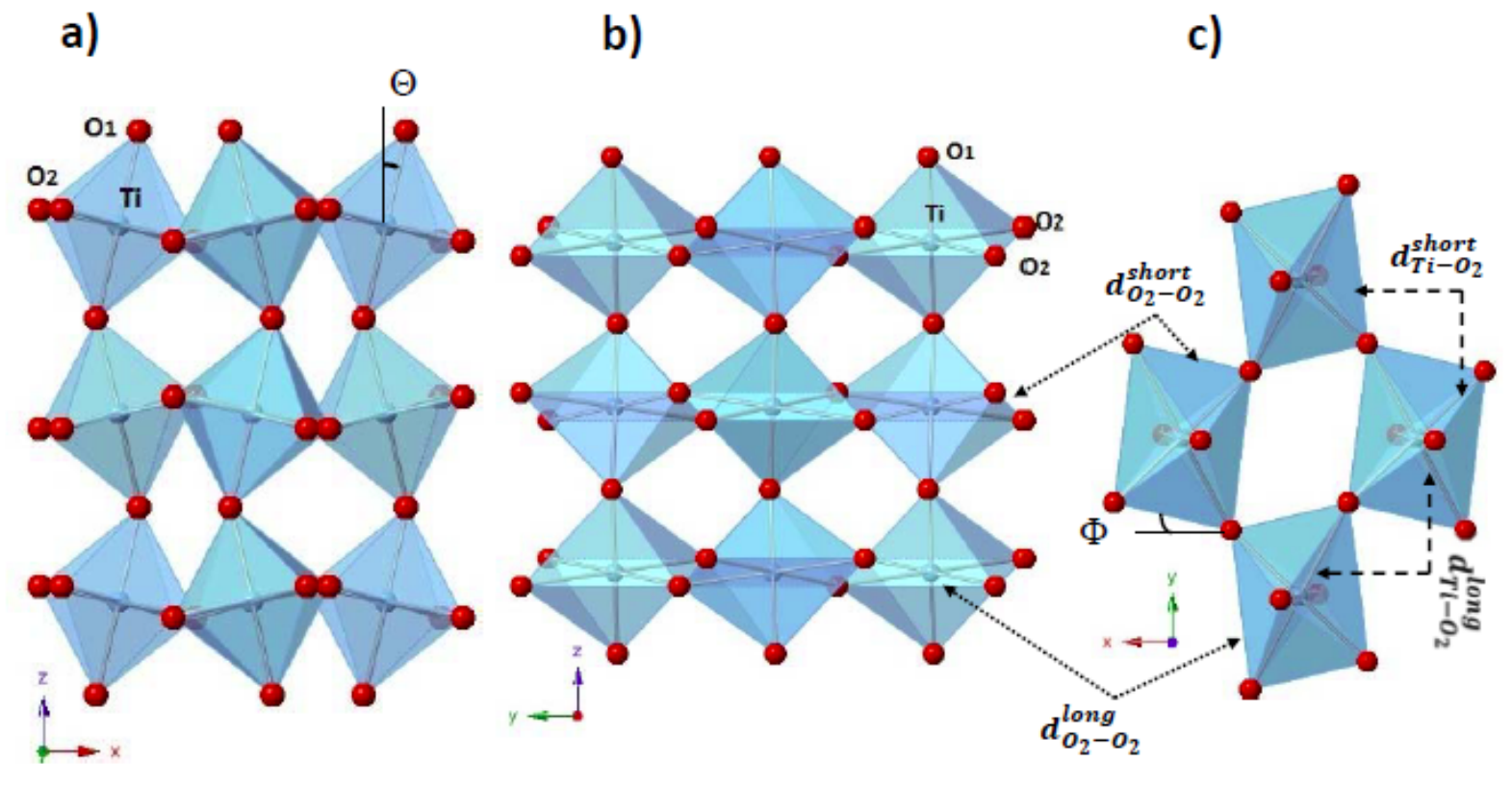}
	\caption{\label{fig:LTO}(Color online) Tilted and rotated \tio octahedra in \LTO~viewed along: \textbf {a)} 010 direction showing the octahedral tilts ($\Theta$) with respect to  the $z$-axis \textbf {b)} 100 direction featuring the rectangular-like basal plane of the \tio octahedra and \textbf {c)} 001 direction  showing clearly the octahedral rotation ($\Phi$), the orthorhombic distortion and the different interatomic distances arising from it. Note that the lattice parameters $b$ have been increased by 35\% for a better  illustration of basal plane distortion.}
\end{figure}

% ------------
\LTO~adopts a $Pbnm$, GdFeO$_{3}$-type orthorhombic structure ($a\ne b\ne c$) with tilted and rotated \tio octahedra (see Figure~\ref{fig:LTO}) in the simulation supercell of 20 atoms, with no experimental evidence of a simpler cubic structure. 
The tolerance factor of a  perovskite compound $ABO_3$, $t$, is defined as the ratio of the intrinsic sizes of the $AO$ square  and the $BO_2$ square:\cite{Goldsmith:1928, Uchida:2003} \begin{equation}
t=\frac{r_A+r_O}{\sqrt{2}(r_B+r_O)}
\end{equation} 
 where $r_A$, $r_B$ and $r_O$ are the ionic radii of each ion. 
In \LTO, the tolerance factor is too small to favor the cubic structure due to the relative size of La$^{+3}$ ion to the \tio octahedra. 

Nevertheless, the octahedral tilts ($\Theta$) stabilize the structure by shortening some La-O bonds (see Figure~\ref{fig:LTO}-a)  and causing Ti-O$_1$-Ti angles to deviate from 180\degree. Any neighboring pair  the \tio octahedra would tilt around the $z$-axis  in  opposite directions.   The subsequent rotation  with respect to the same axis ($\phi$), is restricted to  O$_2$ oxygens forming the basal plane of the \tio and  occur in the same sense. In addition, Cwik et al.\cite{Cwik:2003} showed that the negative sign of their  orthorhombic distortion parameter ($\epsilon=(b-a)/(b+a)$) demonstrates that the \tio octahedra are distorted. The \tio~basal plane is rectangular-like instead of cubic with the O$_2$-O$_2$ edge lengths splitting into $d^{long}_{O_2-O_2}$ and $d^{short}_{O_2-O_2}$ with a ratio:

\begin{equation}
 r_{O_2-O_2}=\frac{d^{long}_{O_2-O_2}}{d^{short}_{O_2-O_2}}
\end{equation} 

 $r_{O_2-O_2}$ reaches up to 1.04 at 8~K. The octahedron's basal plane angles also deviate from the ideal 90\degree  (see Figure~\ref{fig:LTO}-c) leading to  differences in the Ti-O$_2$ distances namely, $d^{long}_{Ti-O_2}$ and $d^{short}_{Ti-O_2}$ and the ratio: 
\begin{equation}
 r_{Ti-O_2}=\frac{d^{long}_{Ti-O_2}}{d^{short}_{Ti-O_2}}
\end{equation} 
\begin{figure}[!htb]
\includegraphics[width=0.48\textwidth]{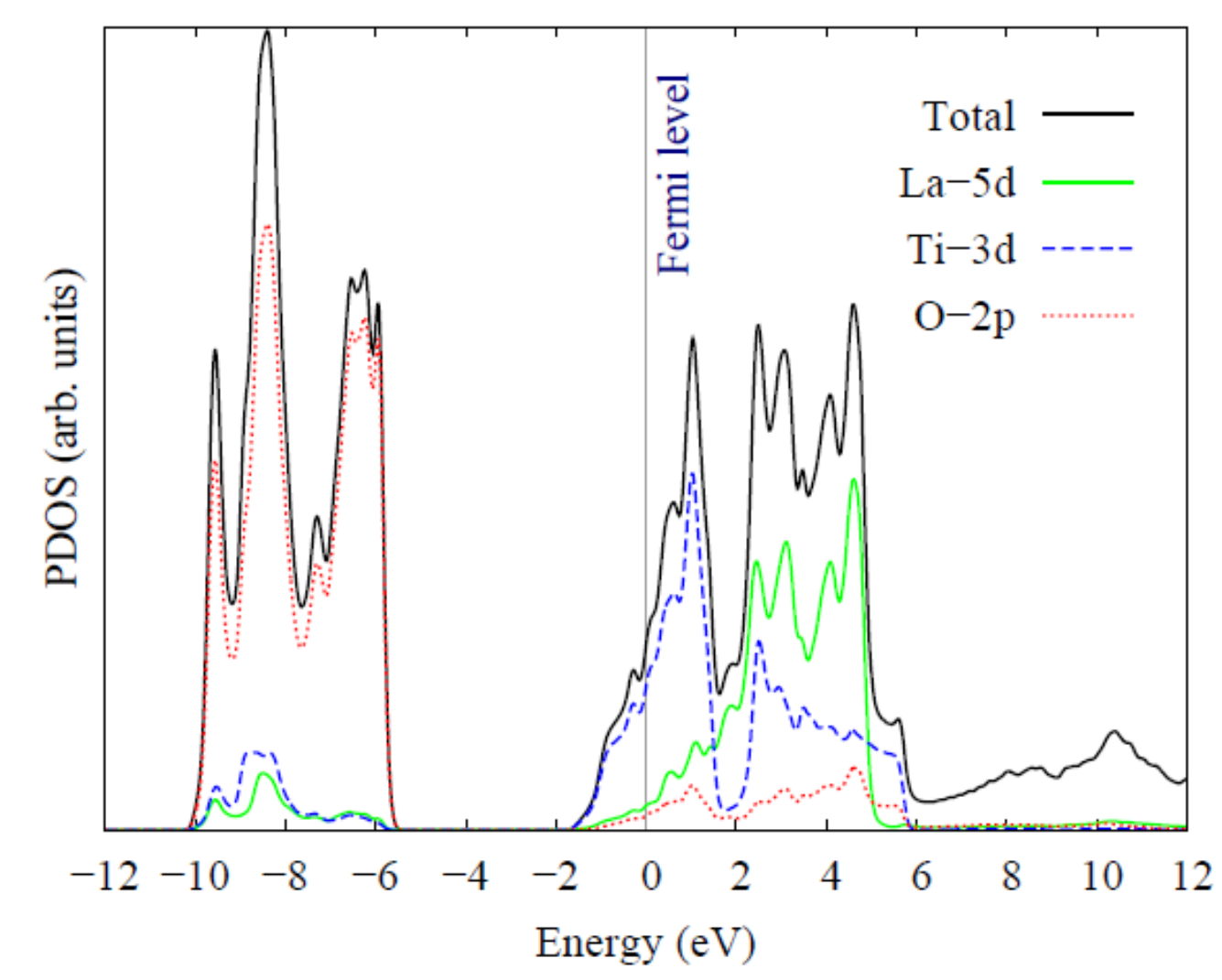}
\includegraphics[width=0.48\textwidth]{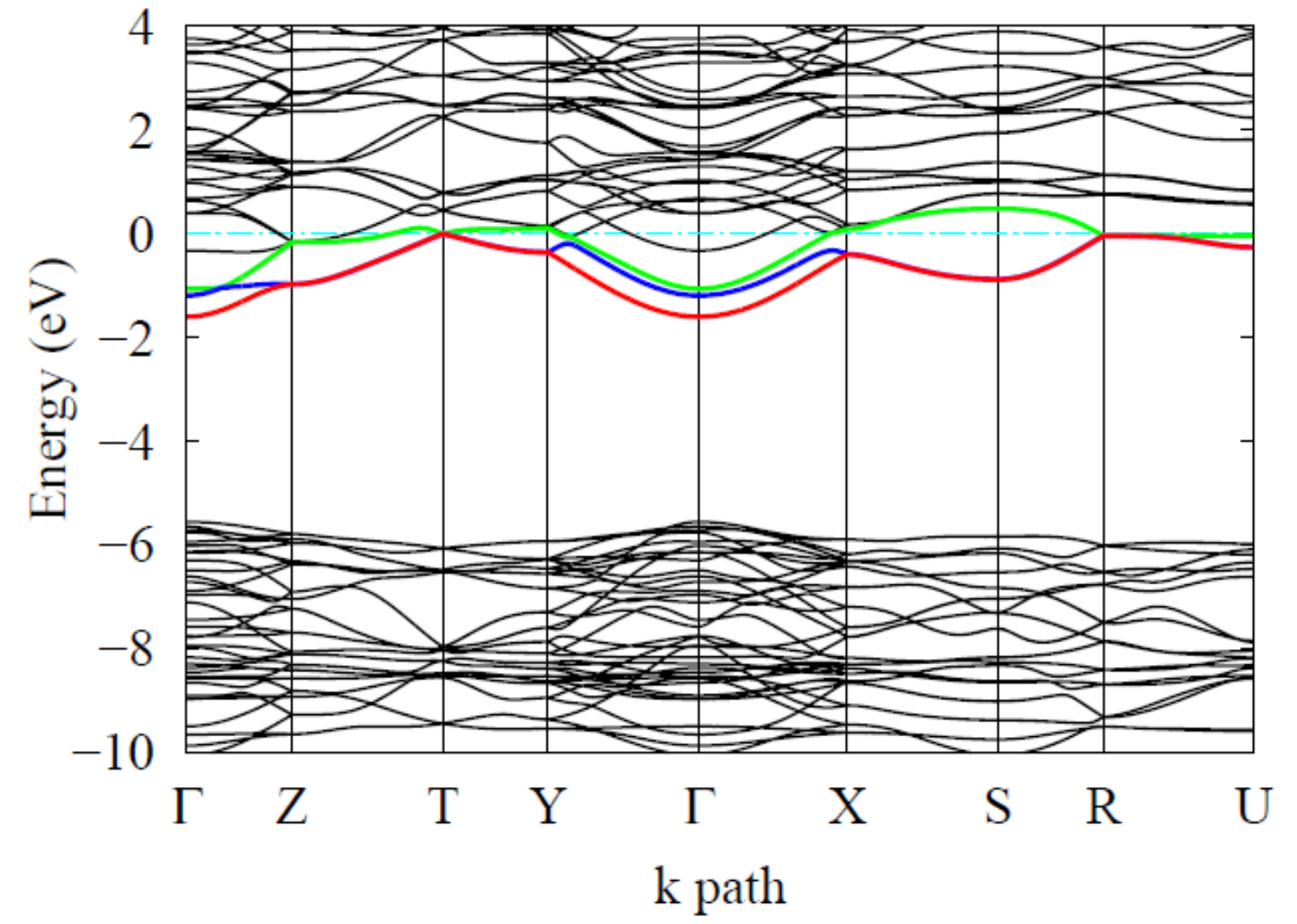}
\caption{\label{fig:pdosBS-HSE}(Color online) \textbf{Top:} Total and projected density of states for orthorhombic  \LTO~as calculated with HSE06 and the SZVP basis set. Only the contribution from the dominant states are represented. \textbf{Bottom:} The corresponding band structure of \LTO~ in the non-magnetic state. The Ti $t_{2g}$ triplet bands can be distinguished by their colors where the splittings caused by octahedral deformation can be clearly seen.} 
\end{figure}

\LTO~undergoes a phase transition from a non-magnetic insulator to an antiferromagnetic Mott insulator at temperatures near T$_c$=146 K without undergoing  important changes in the structural parameters.\cite{Cwik:2003} To determine the initial geometry for LTO, we fully-relaxed a non-magnetic 20-atom LTO supercell with the PBE and HSE06  functionals, starting from several experimental structures acquired at different temperatures.\cite{Cwik:2003,Eitel:1986,Mclean:1979}  
All structures relaxed to the same minimum. 
The lattice parameters of the fully-relaxed non-magnetic structure with both PBE and  HSE06 are summarized in Table ~\ref{tab:LTO_lattice} and compared to experimental data taken at T=155 K just above the phase transition temperature. 
%
%%%
\begin{table*}[!htb]
\caption{Lattice  constants and orthorhombic distortion parameters   as well as octahedral tilt ($\Theta$) and rotation angles ($\Phi$) calculated using SZVP basis set for the fully-relaxed \LTO~ structures. Comparison is done with experimental data taken at  155K from Ref.~\onlinecite{Cwik:2003} near the magnetic to antiforromagnetic transition temperature $T_N$= 146~K. } 
\begin{ruledtabular}
\begin{tabular}{lllllllll}
\label{tab:LTO_lattice}

	&a (\AA)	&b(\AA)   &c(\AA) &$\epsilon$ &$r_{O_2-O_2}$\footnotemark[4]  &$r_{Ti-O_2}$\footnotemark[4] &$\Theta$(\degree) &$\Phi$(\degree)\\
\hline
Experiment\footnotemark[1] 	&5.635	&5.602 &7.905 &-0.0030 &1.035  &1.011 &12.86 &9.69\\
HSE06  &5.600    &5.534    &7.904    &-0.0059 &1.027 & 1.013 &10.88 &9.25\\
PBE      &5.595    &5.632    &7.938    &+0.0033 &1.008 &0.998  &9.54 &6.85\\
LSDA\footnotemark[2]	&5.462	&5.524	&7.789	&+0.0056	&\textendash\ &\textendash\ &\textendash\ &\textendash\\
LSDA$+U$\footnotemark[3] &5.586 &5.529 &7.89 &-0.0051	&\textendash\ &\textendash\ &\textendash &\textendash\  \\

\end{tabular}
\end{ruledtabular}
\footnotetext[1]{At 155K from Ref.~\onlinecite{Cwik:2003}}
\footnotetext[2]{Ref~\onlinecite{Ahn:2006}}
\footnotetext[3]{For the optimum values of $U$=3.2 and $J$=0.9~eV identified in Ref.~\onlinecite{Ahn:2006}}
\footnotetext[4]{Distances evaluated for the same set of reference atoms.}
\end{table*}
%%%

Using HSE06, we observe excellent agreement with  experiment as shown by the orthorhombic distortion parameter\cite{Pickard:2000}  ($\epsilon$) being of the right sign and magnitude. In addition, $r_{O_2-O_2}$ and $r_{Ti-O_2}$ ratios compare very well with experiment\cite{Eitel:1986,Cwik:2003} an indication that the octahedral distortion is well reproduced. 
With PBE, $\epsilon$ has a positive sign  which  suggests that the \tio octahedra  elongate along $b$ instead of $a$ which qualitatively incorrect. 
The last behavior has been reported previously in Ref.~\onlinecite{Ahn:2006}  as a drawback of LSDA that have been overcome  by adding the adequate $U$ correction\cite{Solovyev:1996}  (See the LSDA$+U$ results in Table~\ref{tab:LTO_lattice}). Also, $r_{O_2-O_2}$ and $r_{Ti-O_2}$ ratios indicate that \tio octahedra  are nearly ideal and do not reproduce the distortion observed experimentally. 
The octahedral tilt ($\Theta$) and rotation angles ($\Phi$)  are underestimated by 2 and 0.45\degree,~respectively with HSE06, but the underestimation is worse in the PBE case, reaching 3.3 and 2.85\degree,~respectively.  

Turning to the electronic properties, the calculated projected density of states for \LTO~in the non-magnetic state (Figure~\ref{fig:pdosBS-HSE}) are similar for HSE06 and PBE  showing metallic behavior and a Fermi level lying at the middle of the Ti-3$d$ band.\cite{Okatov:2005}
The  Ti-$3d$ band is separated from the O-$2p$ band by 3.95 and 2.87~eV  for HSE06 and PBE, respectively. 
As with \STO, the valence band (VB) is dominated by O-$2p$ states while the conduction band is dominated by Ti-$3d$ states with some intermixing between O-$2p$ and Ti-$3d$ orbitals, indicating a partially covalent bond.

\begin{figure*}[!htb]
\includegraphics[height=0.25\textheight]{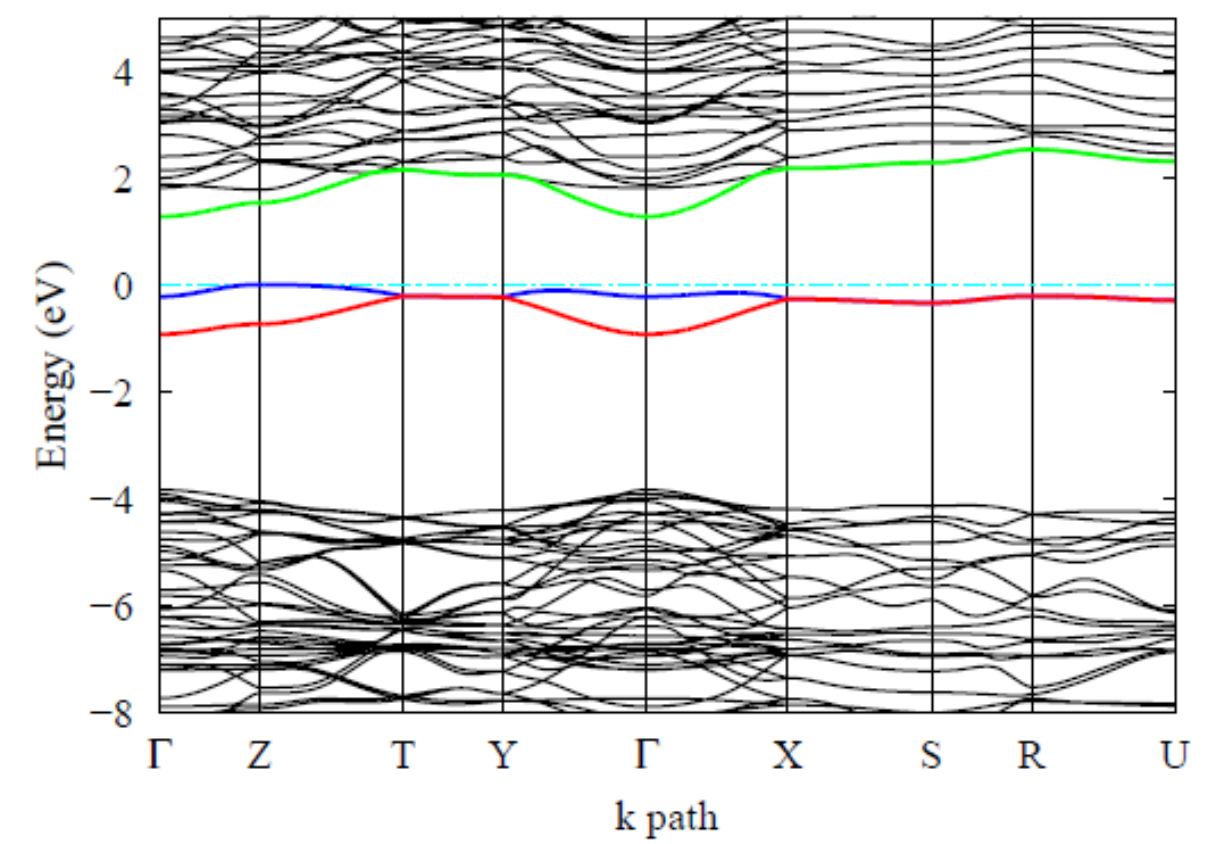}\includegraphics[height=0.25\textheight]{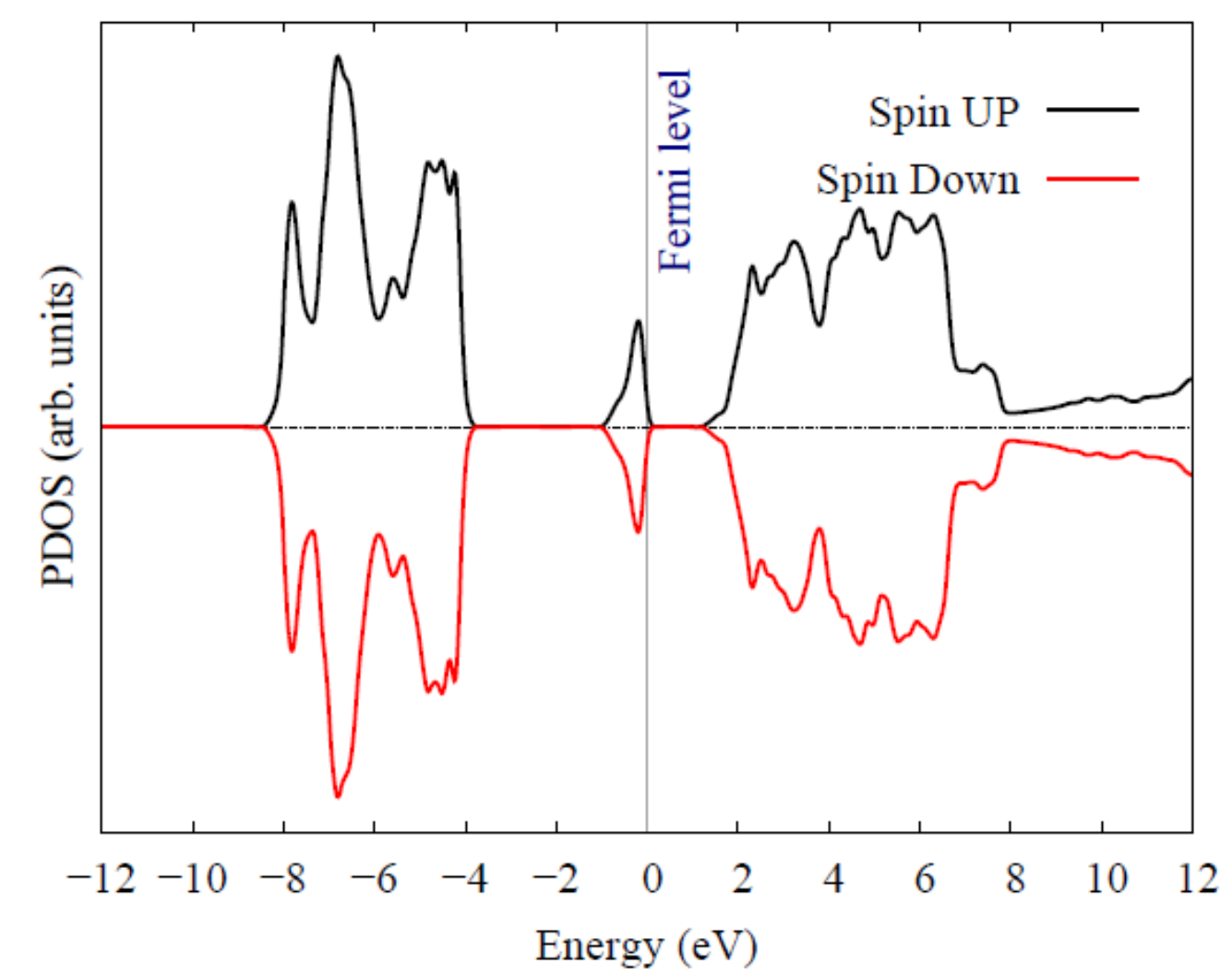}
\caption{\label{fig:LTO_pdos}(Color online) Band structure and PDOS for \LTO~ in the G-Type antiferromagnetic spin orientation.  The Ti $t_{2g}$ triplet bands can be distinguished by their colors and $\Delta^{magn}_{tot}$ is evaluated from the separation of the two bottom ones.} 
\end{figure*}

\begin{figure}[!htb]
\includegraphics[width=0.5\textwidth]{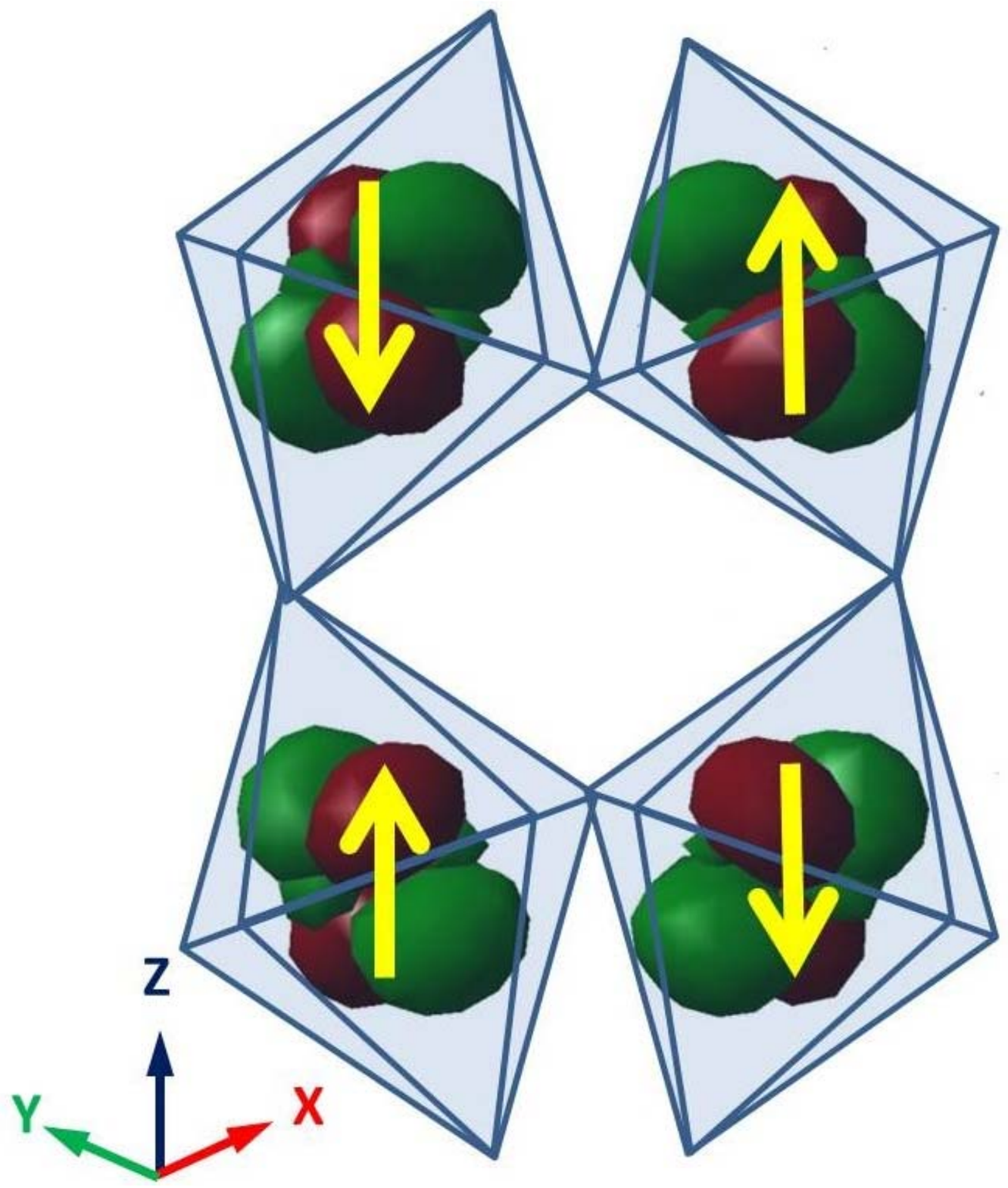}
\caption{\label{fig:orbitals}(Color online)   Isosurface of the highest occupied t$_{2g}$ state  orbitals  for G-type AFM \LTO~at the $\Gamma$ special k point. Orbitals show a ferro-orbital ordering alternating between $d_{xz}$ and $d_{yz}$ orientations between each neighboring \tio octahedra.  Arrows indicate the antiferromagnetic spin orientation on  Ti sublattice.}
\end{figure}
The band structure in the bottom of figure~\ref{fig:pdosBS-HSE} show more details about the O-$2p$ and  Ti-$3d$ band separation where the $t_{2g}$ triplet states\cite{Pavarini:2005, Krivenko:2012} can be distinguished by their colors. The GdFeO$_{3}$-type distortion causes a  large crystal field splittings ($\Delta_{CF}$) defined as the separation  between the two first $t_{2g}$ triplet states\cite{Pavarini:2005, Krivenko:2012} at the $\Gamma$ point and a subsequent smaller splitting ($\Delta_{CF}'$) between the second and the third states.  The amount of $\Delta_{CF}$ is strongly correlated to the orbital state of LTO (see the review paper of Mochizuki and Imada\cite{Mochizuki:2004} and the recent paper of Krivenko\cite{Krivenko:2012}) with small CFS inducing orbital liquid state (OL),\cite{ Solovyev:2004,Solovyev:1996} while  large CFS  favoring an induced orbital ordering (IOO).\cite{Haverkort:2005,Streltsov:2005, Cheng:2008,Filippetti:2011}

 With HSE06, $\Delta_{CF}$ is about 410 meV, somewhat larger than the experimental splittings of 240 meV\cite{Cwik:2003} and 300 meV~\cite{Haverkort:2005} evaluated for non magnetic solutions.  With PBE, we find a much smaller $\Delta_{CF}$ of 61~meV, which is consistent with LSDA$+U$ calculations ranging  between 37-54 meV.\cite{Solovyev:2006,Solovyev:2004} This small CFS is a shared feature between semilocal functionals and  might be at the origin of the unsettled  theoretical description of the orbital state of bulk \LTO. \cite{Mochizuki:2004,Krivenko:2012}  With HSE06, $\Delta_{CF}'$ =190 meV which is high compared to the nearly degenerate doublet observed experimentally.~\cite{Cwik:2003} This difference probably comes from the higher orthorhombic distortion we observe in our  defect-free relaxed structures which is known experimentally to increase with decreasing defect concentration.\cite{Cwik:2003}

%See Table ~\ref{tab:LTO_band} for a summary of $\Delta_{CF}$ values.\cite{Krivenko:2012}

%%%%%
 \begin{table*}[!htb]
\caption{ Calculated HSE06 energy differences referenced to the G-type antiferromagnetic (AFM) ground state ($\Delta$E), the Mott gap ($M_{gap}$), the charge transfer gap ($\Delta$) and the  magnetic moments ($\mu$)  for the Ti atoms  as obtained from the Mulliken population analysis of the spin density. The separation between the O-$2p$ states and the first Ti-$3d$ states ($\Delta_{pd}$), the width of the Mott band ($E_{M}$) as well as the value of the first splitting of the t$_{2g}$ at the $\Gamma$ point ($\Delta^{magn}_{tot}$) are also reported.  } 
\begin{ruledtabular}
\begin{tabular}{lccccccccc}
\label{tab:LTO_band}
		    &&Spin &$\Delta$E(meV) &$M_{gap}$(eV) &$\Delta$(eV) &$\mu$(a.u.) & $\Delta_{pd}$(eV)  & $E_M$(eV) & $\Delta^{magn}_{tot}$(meV)\\
\hline
\\
\multicolumn{2}{l}{AFM: G-Type} &&&&&\\
Experiment 	&& 		&--   &0.1-0.2 &4.5 &0.57\footnotemark[1],0.45\footnotemark[2]	&3.0\footnotemark[2]	& &120-300\footnotemark[3]\\

Prev. Calcs. &&&--&1.6\footnotemark[3], 0.57\footnotemark[4]	&5.2\footnotemark[3], 3.5\footnotemark[4]  
&0.89\footnotemark[3], 0.78\footnotemark[4] &3.1\footnotemark[3], 2.3\footnotemark[4] & &400\footnotemark[3], 230\footnotemark[4] \\	
& & 	&--&0.77\footnotemark[5]	&  &0.68\footnotemark[5] & & &140\footnotemark[6], 54\footnotemark[7] \\
\\		
\hline
\\
This Work &&&&&\\
AFM G-Type && $\begin{array}{cc} \uparrow & \downarrow \\\downarrow& \uparrow  \end{array} $	&0 &1.27 &5.2  &0.94 	&2.9	&1.0	&700\\

AFM A-Type &&$\begin{array}{cc} \downarrow & \downarrow  \\ \uparrow& \uparrow  \end{array} $   &21	&1.39 &5.0 & 0.90 	&3.0	&0.6	&326\\
AFM C-Type &&$\begin{array}{cc} \uparrow & \downarrow  \\ \uparrow& \downarrow  \end{array}$	&44 &0.93 &4.9  &0.92	&2.9	&1.4	&660\\
FM 			&& $ \begin{array}{cc} \uparrow & \uparrow  \\ \uparrow& \uparrow  \end{array} $		&795	   &1.02  &5.2 &0.92 	&2.5	&1.8	&720\\
%Non-Magn	&	&1853		&0	&\textendash &0	& &4.0	&410\footnotemark[4] \\
\end{tabular}
\end{ruledtabular}
\footnotetext[1]{Ref~\onlinecite{Cwik:2003}}
\footnotetext[2]{Ref~\onlinecite{Goral:1983}}
\footnotetext[3]{Ref~\cite{Haverkort:2005}}
\footnotetext[4]{VPSIC results from ref~\onlinecite{Filippetti:2011} for non relaxed strucutres.}
\footnotetext[5]{LDA$+U$ results form ref~\onlinecite{Streltsov:2005} using the experimental \LTO~structure at 8 K from Ref~\onlinecite{Cwik:2003}.}
\footnotetext[6]{DFT$+U$+GWA form ref~\onlinecite{Nohara:2009}.}
\footnotetext[7]{LDA$+U$ results form ref~\onlinecite{Pavarini:2004}.}
\footnotetext[8]{LDA$+U$ results form ref~\onlinecite{Solovyev:2004}.}
\end{table*}
 %%%
% 

Turning now to the properties of magnetic solutions; the experiments of Cwik et {\it al.},~\cite{Cwik:2003} reveal  almost no difference in  the structural properties between the magnetic and non-magnetic solutions close to the transition temperature.  Based on that,  we assume that the   calculated structure in the magnetic state do not differ  much from the non magnetic one near the transition temperature leading to  $\Delta_{CF}^{nonmagn}$=$\Delta_{CF}^{magn}$.   Thus, the  total splitting of the t$_{2g}$ states in a magnetic state ($\Delta^{magn}_{tot}$) is a result of the interplay between $\Delta_{CF}^{nonmagn}$  due to the  structural GdFeO$_3$ distortion and  spin superexchange splitting ($\Delta_{SE}$).
% Ideally, one would like to fully relax the magnetic solutions also, but our calculation did not converge.

The non-magnetic minimum was used for subsequent \textit{unrestricted} spin calculations where spin flips were carried out to simulate, using the same  parameters,  the ferromagnetic (FM) and G-type, A-type and C-type antiferromagnetic (AFM) states (see Table~\ref{tab:LTO_band} and  Ref.~\onlinecite{Goodenough:1955} for a description of the different magnetic orderings). Two reasons are beyond considering the different magnetic states: First, we want to test the ability of HSE06 to identify the  ground state because previous calculation with LDA+$U$,~\cite{Streltsov:2005} predicted the A-type AFM state to be the ground state, which is in disagreement with the G-type AFM state found experimentally. Second, those different magnetic orders might not be observed in bulk LTO under normal conditions, but might become relevant under strain or for  LTO based heterostructures. Thus,  details of electronic structure of LTO in all magnetic states will be examined with focus on $\Delta^{magn}_{tot}$.

In the ferromagnetic  and  antiferromagnetic  spin orientations, PBE predicts LTO as metallic, whereas experiment has shown it to be a Mott insulator.\cite{Cwik:2003} 
This is a well-known limitation of semilocal functionals and agrees very well with the conclusions from early DFT studies  of LTO using planewave basis sets.\cite{Ahn:2006,Okatov:2005,Hamada:1997,Pari:1995} 
This has been overcome in the past by adding the Coulomb term correction ($U$) or via a GW correction.\cite{Solovyev:2008,Nohara:2009} HSE06 performs much better than PBE predicting the G-type AFM  ordering to be the ground state and a Mott insulator.\cite{Cwik:2003}
\LTO~also transforms to a Mott insulator for the other magnetic orders we considered,  with an energy gain relative to ground state of 21 meV  for A-type AFM, 44 meV  for C-type AFM and 795 meV for the FM states. 

The band structure of a Mott insulator is characterized by the optical gap, which is determined by the Hubbard splitting ($M_{gap}$) of the $d$-band separation of the lower and upper Hubbard bands; this is also called the Mott gap. 
The other gap is called the charge transfer gap, which is the energy difference ($\Delta$) between the filled $p$-bands and the unoccupied upper Hubbard $3d$-band. 
For \LTO, the experimental Mott gap is 0.1 eV according to the measurements of Arima \textit{et al.}\cite{Arima:1993} while the CT gap is 4.5 eV. 

The nature of the band gap (direct vs. indirect), the shape and the width of the Mott bands as well as  the amount of $\Delta^{magn}_{tot}$ are discussed below: In the G-type AFM (Figure~\ref{fig:LTO_pdos}),  the gap is indirect from $Z\rightarrow\Gamma$ and measures 1.27~eV, separating the occupied lower Hubbard band from the unoccupied higher Hubbard  band, the Mott band is 1~eV, showing  a $\Delta^{magn}_{tot}$ of 700 meV.  
 In the A-type AFM ordering we observe a larger, indirect band gap from  $R\rightarrow\Gamma$ of 1.39~eV, but a narrower Mott band (sharper and thinner) due to the reduced splitting between the $t_{2g}$ states, ca. 320 meV. 
In the C-type AFM spin orientation, the indirect gap from $S\rightarrow\Gamma$ has the smallest value, 0.9~eV, but  $\Delta^{magn}_{tot}$ is comparable to the G-type ordering value while the Mott band is larger by 0.4~eV. 
In the FM state, we find a direct band gap of 1~eV and $\Delta^{magn}_{tot}$ comparable to the G-type ordering value while the Mott band is larger by 0.8~eV. 

Our calculated Mott gaps, CT gaps and magnetic moments for various magnetic orders are summarized in Table ~\ref{tab:LTO_band} and compared to experiments and previous calculations. The separation between the O-$2p$ states and the first Ti-$3d$ states ($\Delta_{pd}$), the width of the Mott band ($E_{M}$) as well as the value of the first splitting of the t$_{2g}$ at the $\Gamma$ point ($\Delta^{magn}_{tot}$) are also reported. In the G-type AFM ground state, our HSE06 value for $\Delta_{pd}$=2.9 eV is in  excellent agreement with the 3~eV value reported experimentally.~\cite{Goral:1983} Also, the CT gap is only  0.5~eV higher than the  experimentally measured value of 4.5~eV, which constitutes an improvement over the previous LDA$+U$ calculation.\cite{Streltsov:2005}
However, $M_{gap}$ and  $\mu$ are higher than the experimentally reported values. This  overestimation obtained with HSE06 was addressed in the recent study of  Rivero \textit{et al.},\cite{Rivero:2009} which demonstrated that   screened hybrids do provide a good \textit{quantitative} description of the electronic structure of strongly-correlated systems, but that the magnetic coupling constants remain overestimated\cite{Prodan:2007,Prodan:2006,Prodan:2005} compared to experiment. 
This phenomenon is observed using other theoretical methods,\cite{Filippetti:2011,Nohara:2009,Okatov:2005, Streltsov:2005, Pavarini:2004,Solovyev:2004} summarized in Table~\ref{tab:LTO_band} and several explanations for this wide-spread divergence of theory and experiment have been proposed.\cite{Filippetti:2011,Mochizuki:2004} It is very likely that the experimental samples are not pristine and contain defects which ``metallize''them, yielding smaller gaps.

%Mochizuki and Imada~\cite{Mochizuki:2004} posit that the experimentally-measured  insulating and magnetic properties of \LTO~samples are very sensitive to slightly off-stoichiometry compositions, unlike the defect-free structures used in calculations, and that the higher defect concentration  leads to reduced magnetic moments and band gaps. 
%In the case of actinide oxide Mott systems, HSE06 performs rather well,\cite{Prodan:2007,Prodan:2006,Prodan:2005} yet the Mott gaps and magnetic moments predicted for \LTO~ are higher than the measured values.

Based on our HSE06  large $\Delta_{CF}$ and $\mu$ values, LTO is expected to display an   induced orbital order (IOO). Indeed, orbital ordering in  G-type AFM \LTO~can be clearly seen  in Figure~\ref{fig:orbitals} showing the highest occupied t$_{2g}$ state  orbitals  for G-type AFM \LTO~at the $\Gamma$ special k point. Orbitals show a ferro-orbital ordering oriented parallel to the $z$-axis and oriented along the long Ti-O$_2$ bonds. Because  the {\it long} Ti-O$_2$ bonds (see Figure~\ref{fig:LTO}-c) are perpendicular between each neighboring \tio octahedra,  the orbitals sitting along those bonds appear alternating between $d_{xz}$ and $d_{yz}$ and  tilted due to the GdFeO$_3$ distortion.  This picture confirm once again  experiments\cite{Haverkort:2005} and theoretical calculations\cite{Mochizuki:2004} suggesting that  \LTO~has an  induced orbital order and might not be an orbital liquid.  
 
% because this have been recently published using VPSIC method by Filippetti et al.~\cite{Filippetti:2011} but in this paper the supercell parameters were fixed to the experimental values at 8K. Our approach is different, as we focus on a full relaxation of the structure giving reasonably good structural parameters with HSE and trustable crystal field splitting that compare better with experiment than PBE and reveal that LTO favors orbital ordering. 

%Recently,  Krivenko.~\cite{Krivenko:2012} proposed a  unified  theory  for $d^1$ perovskites defining a critical value $\Delta^*$=4/9$J_{SE}$ separating the Induced Orbital Order (IOO) regime ($\Delta_{CF} > \Delta^*$) and Orbital Liquid (OL) regimes ($\Delta_{CF} <\Delta^*$) in terms of CF and superexchange parameters. Using the calculated total splitting  $\Delta_{tot}^{magn}$=700~meV in the G-type AFM \LTO we estimate  $\Delta_{SE}$ = 290~meV and  Kirivenko  critical value   $\Delta^*$=129 meV. 
% ====
\section{Conclusions}
% ====
The performance of the screened hybrid HSE06 was tested for modeling  the fully relaxed  semi-conducting metal oxide perovskites \STO~and \LAO~in their different structural phases, and  the strongly-correlated Mott insulator \LTO~in the ferromagnetic and various antiferromagnetic orderings. HSE06 gives good band gaps and accurate structural/geometric properties. 
It  demonstrated great  efficiency in handling structural phase transitions,  and yielded accurate structural properties and order parameters (octahedral rotations/tilts and strains) for all three perovskites.  
This is a substantial improvement over DFT+$U$ -- one that suggests that HSE06 may be as successful in treating other metal oxides and  the metal-insulator transition  in oxide superlattices. 
In addition, the crystal field splitting ($\Delta_{CF}$) of the $t_{2g}$ states resulting from the phase transitions in  \STO, \LAO~and \LTO~was evaluated for all three materials, and showed excellent agreement with experiment in the STO case.\cite{Uwe:1985} The LTO computed $\Delta_{CF}$ = 410~meV is rather high and indicate that \LTO~has an induced orbital order that shows clearly from our analysis of highest occupied t$_{2g}$ state  orbitals.

%: ||||||||||||||||||||||||||||||||||||   ACKNOWLEDGMENTS    |||||||||||||||||||||||||||||||||||||||||||||||||||||||| 
\begin{acknowledgments}

This work is supported by the  Qatar National Research Fund  (QNRF) through the National Priorities  Research Program (NPRP  08 - 431 - 1 - 076). We are grateful to the Research computing facilities at Texas A\&M University at Qatar for generous allocations of computer resources.\\

\end{acknowledgments}

\end{document}